\documentclass[aps,prl, twocolumn,superscriptaddress,longbibliography]{revtex4-2}
\usepackage{amsmath,amssymb,amsfonts,amsthm}
\usepackage{color}
\usepackage{newtxtext}
\usepackage{newtxmath}
\usepackage[colorlinks,citecolor=blue,linkcolor=blue,urlcolor=blue]{hyperref}
\usepackage{verbatim}
\usepackage{graphicx}
\usepackage{dblfloatfix}

\usepackage[ruled, vlined]{algorithm2e}
\SetKwInOut{Input}{Input}
\SetKwInOut{Output}{Output}

\begin{document}

\title{Generalized Reduced-Density-Matrix Quantum Monte Carlo Gives Access to More}
\author{Zhiyan Wang}
\affiliation{State Key Laboratory of Surface Physics and Department of Physics, Fudan University, Shanghai 200438, China}
\affiliation{Department of Physics, School of Science and Research Center for Industries of the Future, Westlake University, Hangzhou 310030,  China}
\affiliation{Institute of Natural Sciences, Westlake Institute for Advanced Study, Hangzhou 310024, China}

\author{Zhe Wang}
\affiliation{Department of Physics, School of Science and Research Center for Industries of the Future, Westlake University, Hangzhou 310030,  China}
\affiliation{Institute of Natural Sciences, Westlake Institute for Advanced Study, Hangzhou 310024, China}

\author{Bin-Bin Mao}
\affiliation{School of Foundational Education, University of Health and Rehabilitation Sciences, Qingdao 266000, China}

\author{Zheng Yan}
\email{zhengyan@westlake.edu.cn}
\affiliation{Department of Physics, School of Science and Research Center for Industries of the Future, Westlake University, Hangzhou 310030,  China}
\affiliation{Institute of Natural Sciences, Westlake Institute for Advanced Study, Hangzhou 310024, China}

\begin{abstract}
For a long time, people have been focusing on how to extract more information, such as off-diagonal observables, from the quantum Monte Carlo (QMC) simulation of the partition function, but there have been numerous difficulties, and many of them are insurmountable. In this article, we point out that all the difficulties stem from the starting point of the simulation: calculating a partition function.
We introduce a paradigm shift: when we transform the simulated object from a partition function to a generalized reduced density matrix (GRDM), the difficult problem of measurement can be readily solved.
By designing the GRDM, both equal-time and nonequal-time off-diagonal observables have been measured easily in QMC with a polynomial computation complexity. As a demonstration, the GRDM enables direct access to nonequal-time correlators for dynamical spectra as well as R\'enyi-1 correlators that reveal strong-to-weak symmetry breaking in the mixed state, capabilities that lie beyond the reach of prior methods. This establishes a unified framework for holographic characterization within QMC.
\end{abstract}

\date{\today}

\maketitle
\textit{\color{blue}Introduction.---} 
Quantum Monte Carlo (QMC) provides a promising numerical framework for overcoming the ``exponential wall'' of quantum many-body systems~\cite{Handscomb_1962,Blankenbecler1981Monte, Scalapino1981Monte, Hirsch1982, Hirsch1983Discrete, ceperley1986quantum,sandvik1991quantum, Sandvik1999, Foulkes_RMP2001, SandvikIsing, Evertz2003loop, Assaad_book2008, Sandvik2010Computational, Melko2013Stochastic, Carlson_rmp2015, gubernatis2016quantum, Yan2019, ZY2020improved}. A central limitation, however, is that the class of observables directly accessible in standard path-integral and stochastic series expansion (SSE) formulations is restricted by the sampling basis. For example, general off-diagonal operators that do not appear explicitly in the Hamiltonian are often not directly measurable within the native configuration space~\cite{sandvik1992generalization,yan2023emergent,ZYan2022}. Among the past decades, a lot of efforts have been tried to overcome this challenge: Worm-like algorithms~\cite{NProkofev1998,DL1_OlavF_2002, Dorneich2001, alet2005generalized} offer a direct scheme for measuring two-point Green's functions, yet their extension to arbitrary models and general off-diagonal operators still faces a substantial gap~\cite{NProkofev1999,Dorneich2001,DL2_OlavF_2003,alet2005generalized,wenjing2021measuring,multiDL2025}. Bell-QMC has introduced a two-copy Bell-sampling framework within SSE, enabling unbiased estimation of all Pauli operators in a single simulation~\cite{ymding2025bell}, while its construction is tied to qubit/Pauli structures and Bell-basis sampling. More recently, the bipartite reweight-annealing method has opened a route, in principle, to arbitrary many-body and imaginary-time off-diagonal observables~\cite{BRAZhiyan}, but it relies on a high computational cost to anneal from a known reference point, while such a reference point is not always naturally available. In conclusion, extracting universal off-diagonal information is currently not a particularly effective approach for QMC.

\begin{figure}[t]
\centering
\includegraphics[width=\linewidth]{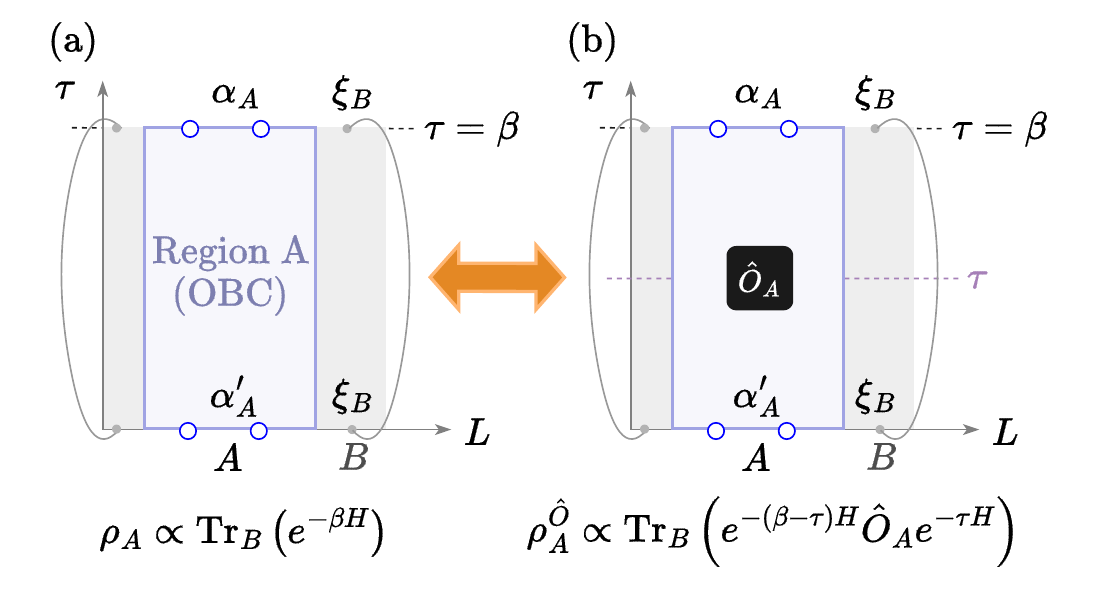}
\caption{Schematic illustration of the reduced density matrix (RDM) and operator-inserted generalized RDM on the A-OBC/B-PBC manifold.
(a) The RDM of subsystem $A$ is obtained by tracing out subsystem $B$, yielding $\rho_A \propto {\rm Tr}_B(e^{-\beta H})$.  Region $A$ (colored) is open in imaginary time, with boundary states $\alpha_A',\alpha_A$ at $\tau=0,\beta$, while region $B$ (gray) remains periodic and is traced over through the states $\xi_B$.
(b) Generalized RDM with an operator insertion $\hat O_A$ at imaginary time $\tau$ inside subsystem $A$, $\rho_A^{\hat O} \propto {\rm Tr}_B \left(e^{-(\beta-\tau)H}\hat O_A e^{-\tau H}\right)$.
The two panels are connected (yellow arrow) by the switch between the inserted operator $\hat{O}_A$ and the identity operator.
}
\label{Fig1}
\end{figure}

Recently, the reduced-density-matrix (RDM) sampling approach proposed by the authors has shown great potential for measuring arbitrary operators in QMC~\cite{Mao2025}. It provides a powerful approach to circumvent the exponential growth of the full Hilbert space~\cite{mao2025detecting,wang2025entanglement,Fabien2025CDM}. In this way, the exponential computational cost is confined to the subsystem $A$, while the degrees of freedom in the environment $B$ are traced out by QMC with polynomial computational complexity.
As is well known, any linear observable $\langle O_A\rangle$ defined on subsystem $A$ satisfies the relation $\langle O_A\rangle=\mathrm{Tr}(\rho O_A)=\mathrm{Tr}(\rho_A O_A)$. This means that measuring an operator $O_A$ in the full system is equivalent to measuring it in the reduced density matrix $\rho_A$ alone. Although the dimension of $\rho_A$ still grows exponentially with the size of $A$, it is sufficient to capture local observables. Therefore, when the support of $O_A$ is fixed, for example for a few-point correlator, the method offers a clear computational advantage in large-scale numerical calculations.  

Due to the significant potential of the sampling RDMs in the field of quantum many-body computation, it has been widely applied in exploring quantum entanglement \cite{mao2025detecting,wang2025entanglement,Mengziyang2025b,Jiang2026indentifying}, magic \cite{Timsina2025robustness} and leading correlation functions \cite{Fabien2025CDM} shortly after its original proposal~\cite{Mao2025}. 
However, to realize this potential more generally, two fundamental challenges must still be addressed:

1) All the measures are limited in the equal-time or static, since a standard RDM contains no explicit dynamical information.  Moreover, once an operator is forcibly inserted at a given imaginary-time point, the basic normalization condition, e.g. $\mathrm{Tr}(\rho_A)=1$, becomes no longer ensured directly.

2) A broadly applicable algorithmic framework for RDM sampling remains absent while the system loses the flippable symmetry of the cluster/loop update~\footnote{Because all the systems studied so far are based cluster or operator-loop scheme of SSE with fixed update-configurations due to a good symmetry, such as TFIMs and Heisenberg models, while the update scheme is uncertain with a probability-based update-path (e.g., XXZ models), the previous methods lose effectiveness.}, i.e., the update is not fixed for a given sampled configuration.

The details of these hardcore problems will be introduced in the following main text. Overall, although the RDM-sampling method displays great potential for extracting information from QMC simulations, it is difficult to extend to general update schemes and nonequal-time measurements.
In this work, we propose the generalized reduced density matrix (GRDM) formulation to overcome both obstacles within a unified framework successfully. The key insight is to extend the conventional reduced description so that reduced information and imaginary-time structure are sampled simultaneously in a single Monte Carlo ensemble. This is achieved through two innovations: (i) a boundary-hole trick, which enables direct RDM sampling within general update schemes such as the SSE directed-loop update and thereby extends the applicability to broader models and parameter regimes; and (ii) a systematic construction of reduced quantities with explicit imaginary-time dependence, which connects RDMs to dynamical information and yields efficient estimators for nontrivial imaginary-time observables. Within the GRDM framework, off-diagonal equal/nonequal correlation functions can be evaluated under the same algorithmic structure, without relying on separate model-specific measurement tricks. In addition, the scheme provides a direct and polynomial-cost estimator for the R\'enyi-1 correlator~\cite{SWSSBZack,SWSSB_Wightman}, which diagnoses symmetry-breaking properties beyond conventional equal-time correlators~\cite{SWSSB_PRXQuantum}. Given this common structure, the formulation applies to both SSE and wordl-line/path-integral QMCs~\cite{Troyer2003nonlocalupdate,Assaad_book2008}, in this work, we use SSE as the main example for concreteness.

\textit{\color{blue}Reduced density matrices.---}
Here we briefly review the basic formulation of the reduced density matrix (RDM). 
The RDM of a spatial subregion $A$ is obtained by tracing out its complement $B$, is given by:
\begin{equation}
 \rho_A = \frac{1}{Z}\mathrm{Tr}_B\bigl( e^{-\beta H} \bigr)
\end{equation}
where $H$ is the Hamiltonian, $\beta = 1/T$ is the inverse temperature, ${Z}=\mathrm{Tr}\left(e^{-\beta H}\right)$ is the partition function as a normalization factor to guarantee $\mathrm{Tr}\rho_A=1$. Expectation values of local operators within region $A$ are obtained from $\rho_A$ as $\langle O_A \rangle = \mathrm{Tr}\left(\rho_A\, O_A\right)$. For example, $O_A$ may represent an equal-time two-point correlation function such as $S_i^x S_j^x$ with $i,j\in A$, giving $\langle S_i^x S_j^x \rangle = \mathrm{Tr} \left(\rho_A\, S_i^x S_j^x\right)$.

In the path-integral QMC or SSE framework~\cite{NProkofev1998,ceperley1986quantum,Deng2002Cluster,Huang2020Worm,sandvik1991quantum,Sandvik1999,Sandvik2010Computational}, the RDM elements $\langle \alpha_A | \rho_A | \alpha_A' \rangle$ is sampled in a spacetime manifold with open boundary condition (OBC) of the $A$ region in the imaginary time direction while the imaginary-time boundary condition of $B$ is still periodic (PBC)~\cite{Mao2025}:
\begin{equation} 
 \langle \alpha_A|\rho_A|\alpha'_A\rangle \propto \frac{\sum_{\{\xi_B\}}\langle \alpha_A, \xi_B | e^{-\beta H} | \alpha_A', \xi_B \rangle}{\sum_{\{\alpha_A,\alpha_A',\xi_B\}}\langle \alpha_A, \xi_B | e^{-\beta H} | \alpha_A', \xi_B \rangle},
 \label{eqrhoa}
\end{equation}
where $\alpha_A/\alpha_A'$ is the bra/ket configuration of $A$ and $\xi_B$ is the configuration in $B$. The QMC illustration of  Eq.\eqref{eqrhoa} is displayed in the Fig.\ref{Fig1} (a) intuitively. Since the imaginary-time direction is OBC in $A$, the bra/ket configurations $\alpha_A,\alpha_A'$ can be obtained in the upper/bottom boundary. Meanwhile, the PBC of $B$ guarantees the bra and ket are the same configuration $\xi_B$. We refer to this hybrid structure as the {A-OBC/B-PBC manifold}.

{\color{blue}\textit{The boundary-hole trick.--}}
To illustrate the essence of the method, we consider the $S=1/2$ XXZ model, simulated by the directed-loop algorithm~\cite{DL1_OlavF_2002, DL2_OlavF_2003}, as a representative example beyond fixed-loop updates such as the operator-loop update in Heisenberg models~\cite{Sandvik1999} and the cluster update in TFIMs~\cite{Sandvik2003Stochastic}. The Hamiltonian is
\begin{equation}
  H = J \sum_{\langle i,j \rangle} \left( S_i^x S_j^x + S_i^y S_j^y + \Delta\, S_i^z S_j^z \right)
  \label{eq:XXZmodel}
\end{equation}
where the summation $\sum_{\langle i,j \rangle} $ runs over nearest-neighbor bonds and $\Delta$ controls the Ising-coupling strength, and $J>0$ is assumed.

The hybrid A-OBC/B-PBC manifold necessitates a refined directed-loop scheme to ensure sampling efficiency. Simply allowing the update line to stop at the open imaginary-time boundary of region $A$, as in conventional operator-loop and cluster updates~\cite{Mao2025,mao2025detecting,wang2025entanglement,Fabien2025CDM}, leads to incorrect results. The reason is that the path-selection at open boundaries contributes an extra probability factor in the detailed balance for uncertain update. 
Results of the drift compared with exact diagonalization (ED), together with the discussion of the detailed-balance violation, are presented in Appendix B.
For this reason, a systematic extension of the update-line scheme remains essential.

To overcome this problem, we introduce a \textbf{\textit{boundary-hole}} trick that embeds the open imaginary-time boundaries of region $A$ directly into the update-line scheme. The key idea is that 
once the update line touches an open boundary in imaginary time, it can be teleported to another open-boundary position. This can be viewed as if there are ``holes'' connected to one another on the open boundary,  providing escape routes for the update lines. In this way, the update line does not stop at the boundary,  but instead continues through the holes until it returns to its starting point and closes, thereby recovering the original directed-loop framework.
According to detailed balance, the selection probability from hole $h$ to hole $h'$ should be equal to that from $h'$ to $h$. For convenience, we therefore choose a uniform probability for teleportation among all holes.

Physically, this mechanism restores the closed topology of the update by allowing loops to tunnel through an additional scattering channel. The benchmark results and technical details are presented in Appendix B. As a result, the RDM can be extracted within general update-lines scheme, and equal-time observables become accessible. This idea also works for cluster-like update, since the update-line scheme is a basic framework in QMC.

\begin{figure}
    \centering
    \includegraphics[width=\linewidth]{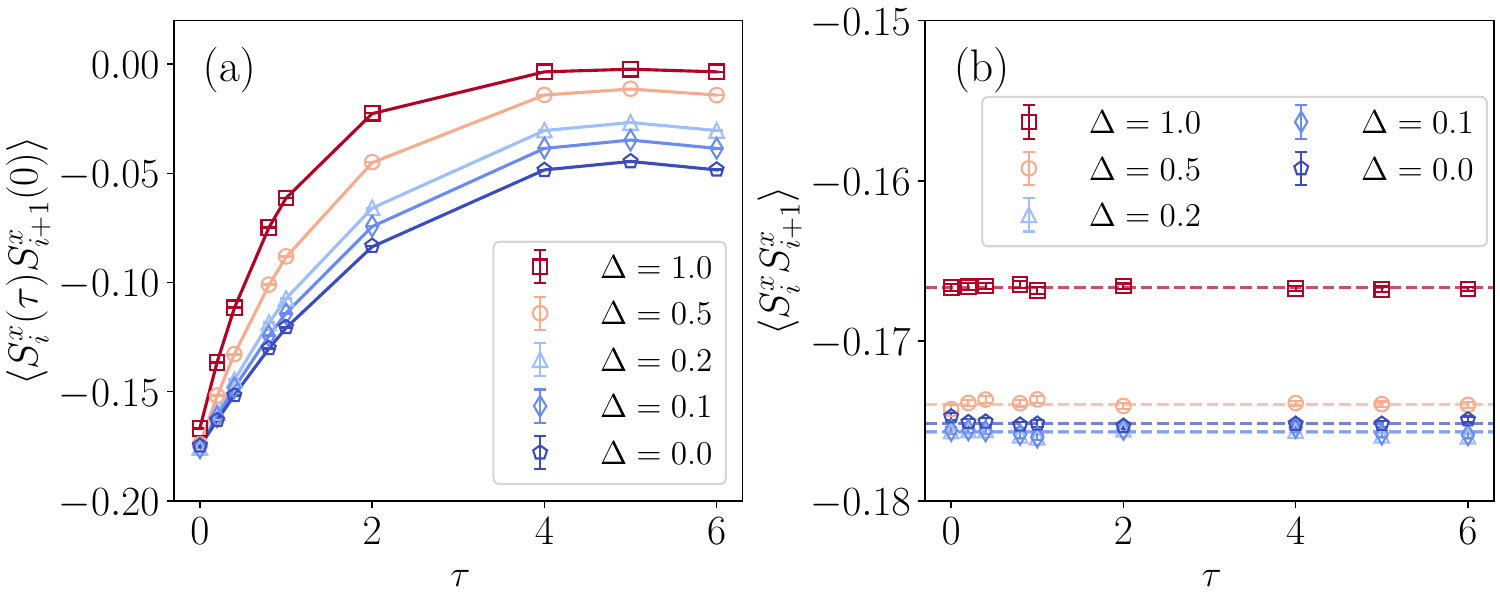}
    \caption{Benchmark results for the $L=4$ XXZ chain with $\beta=10$. The hollow markers (SSE) are consistent with the ED results (dashed lines) within error bars. (a) Imaginary-time correlation function $\langle S^x_i(\tau) S^x_{i+1}(0) \rangle$ as a function of $\tau$ for various anisotropy values $\Delta$. (b) Equal-time correlation function $\langle S^x_i(0) S^x_{i+1}(0) \rangle$ calculated using $\tilde{\rho}^I_A$ from the GRDM framework.}
    \label{fig:imcorbenchmark}
\end{figure}

\textit{{\color{blue}GRDM.--}} While the standard RDM $\rho_A$ captures equal-time measures, it remains a static tomography of the subsystem, and the information regarding the dynamics is typically lost. In order to obtain the dynamical observables, we forcibly insert an operator $\hat{O}_A$ into the RDM at a certain time point $\tau$, and name it as a \textbf{generalized reduced density matrix (GRDM)}. The GRDM is defined as:
\begin{equation}
{\rho}_A^{\hat{O}}(\tau) = \frac{\mathrm{Tr}_B(e^{-(\beta-\tau) H} \hat{O}_A e^{-\tau H})}{Z} 
\label{rhoao}
\end{equation}
This operator corresponds intuitively to inserting the observable $\hat {O}_A$ within region $A$ at imaginary time $\tau$, as illustrated in Fig.~\ref{Fig1}(b). Here $Z=\mathrm{Tr}(e^{-\beta H})$ ensures $\mathrm{Tr}\rho_A=1$, whereas in general $\mathrm{Tr}\rho_A^{\hat{O}}(\tau)\neq1$.

In a Monte Carlo context, normalizing a sampled RDM without insertion in Eq. \eqref{eqrhoa} is straightforward. For the standard RDM, each matrix element is proportional to the sampling frequency of the corresponding boundary configurations, so the sampled matrix can be normalized by forcing the sum of the diagonal ($\alpha_A =\alpha_A '$) frequencies being $1$~\cite{Mao2025}.
However, this normalization method fails when we sample the GRDM of Eq.\eqref{rhoao}, 
because simulating weight proportional to $e^{-(\beta-\tau)H}\hat{O}_A e^{-\tau H}$ on the A-OBC/B-PBC manifold does not by itself produce the correct normalization factor $Z$.

The solution comes from sampling a joint distribution of the GRDM and RDM. Note that when the inserted operator $\hat{O}$ in the GRDM is chosen to be the identity operator $\hat{I}$, the GRDM reduces to the RDM.
Therefore, if we design an effective update scheme between $\hat{O} \leftrightarrow \hat{I}$ in sampling the spacetime manifold, the joint distribution of the GRDM and RDM has been simulated simultaneously by QMC. Since the normalization of the RDM is known from $\mathrm{Tr}\rho_A=1$, this known normalization can be used to fix that of the GRDM as well.  
In practice, we denote the unnormalized GRDM directly sampled in QMC by $\tilde{\rho}_A^{\hat{O}}(\tau)$, and the identity-inserted case by $\tilde{\rho}_A^{\hat{I}}=\tilde{\rho}_A$, namely the unnormalized RDM, such that $\tilde{\rho}_A^{\hat{I}}/Z=\rho_A$ and $\tilde{\rho}_A^{\hat{O}}(\tau)/Z=\rho_A^{\hat{O}}(\tau)$ as in Eq.\eqref{rhoao}.
In this framework,  $\tilde{\rho}_A^{\hat{O}}(\tau)$ and $\tilde{\rho}_A^{I}$ form the numerator and denominator, respectively, of the target measurement in SSE. 
For example, for an inserted operator $\hat{O}^{(1)}_A$ at $\tau$ and another local operator $\hat{O}_A^{(2)}$, the general form of the imaginary-time correlation function is given by:
$\langle O_1(\tau)O_2(0)\rangle=
\mathrm{Tr} \left[\tilde{\rho}^{\hat{O}^{(1)}}_A(\tau)\hat{O}_A^{(2)}\right] / 
\mathrm{Tr} \left[\tilde{\rho}^{\hat{I}}_A \right]$.
The operation of trace will be done additionally after sampling the matrices via QMC to finish the measurement.

The remaining question is how to realize an efficient $\hat{O}\leftrightarrow I$ transition while preserving detailed balance. 
As a specific example, we consider the SSE directed-loop update with $S^x=\frac{1}{2}\sigma^x$ as the inserted operator. 
While such between diagonal and off-diagonal conversion is incorporated in TFIM-type updates~\cite{Sandvik2003Stochastic,Melko2013Stochastic},  
in more general cases, however, the inserted $S^x$ acts as a local defect in the time boundary which is forbidden in the spacetime manifold.  
To maintain the integrity of the directed-loop framework during the transition between operators $I$ and $S^x$, we extend the boundary-hole concept to internal imaginary-time coordinates by introducing a pair of \textit{operator-holes} around the inserted operator, which serve as localized sources and sinks that convert world-line discontinuities into additional scattering channels.  In Appendix~C, we provide a detailed description of the operator-holes construction and its technical implementation.

To benchmark the performance of the GRDM framework, we investigate the transverse correlation functions of the small system, using exact diagonalization (ED) as a numerical reference. Within the GRDM scheme, by employing the operator transition $S^x_i \leftrightarrow \hat{I}$ at site $i$ where $i,j \in A$, we simultaneously construct $\tilde{\rho}_A^{S^x_i}(\tau)$ and $\tilde{\rho}^I_A$. The imaginary-time correlation function is then evaluated via the ratio $\langle S_i^x(\tau) S_{j}^x(0) \rangle = \mathrm{Tr}[\tilde{\rho}_A^{S^x_i}(\tau) S^x_{j}] / \mathrm{Tr}[\tilde{\rho}^I_A]$. \ Meanwhile, the equal-time correlation $\langle S_i^x S_{j}^x \rangle$ can be directly extracted from the trace of $\tilde{\rho}^I_A$ within the same sampling procedure, i.e, $\langle S_i^x S_{j}^x \rangle = \text{Tr}[\tilde{\rho}_A^{I} S^x_{i} S^x_{j}] / \text{Tr}[\tilde{\rho}^I_A]$. As illustrated in Fig.~\ref{fig:imcorbenchmark}, panel (a) shows that the imaginary-time correlation functions reconstructed from the GRDM are in excellent agreement with the ED benchmarks across the entire range of $\tau$ and anisotropy $\Delta$.  Panel (b) shows that the simultaneously sampled $\tilde{\rho}_A^{I}$ can be used as the RDM to evaluate equal-time observables.~\footnote{Here the horizontal axis is still labeled by $\tau$ to indicate that each $\tilde{\rho}_A^{I}$ is obtained together with the corresponding operator insertion at imaginary time $\tau$.  As expected, the resulting equal-time correlation functions show that the sampled $\tilde{\rho}_A^{I}$ should remain no dependence on the operator-insertion time $\tau$.} 
Thus, for the series of independently simulated insertion times used to evaluate the imaginary-time correlator, we simultaneously obtain a corresponding batch of RDM samples.

\begin{figure}[!b]
    \centering
    \includegraphics[width=\linewidth]{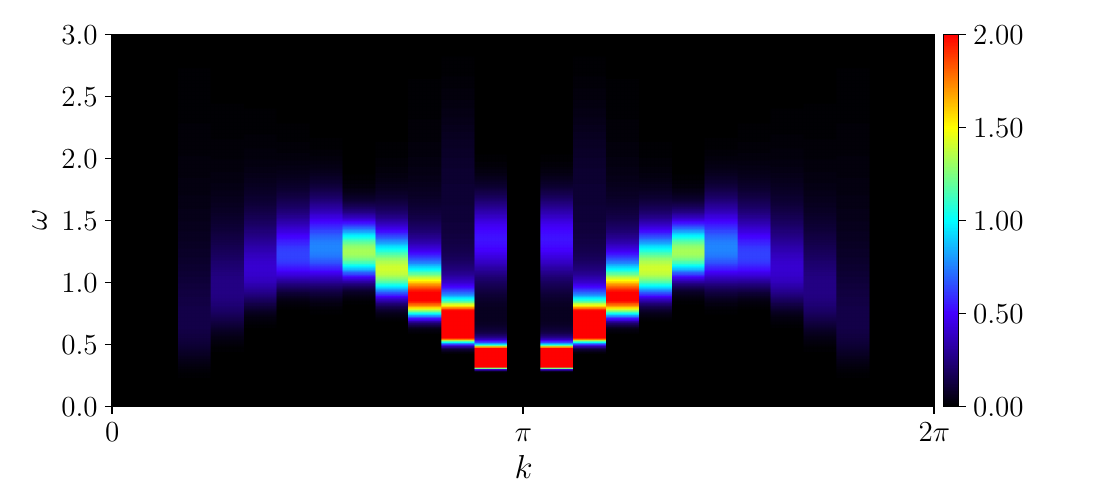}
    \caption{Spectral function $S^{xx}(k,\omega)$ for the spin-1/2 XXZ chain at $\Delta=0.2$. }
    \label{fig:xxspectrum}
\end{figure}

\onecolumngrid

\refstepcounter{figure}
\begin{center}
    \includegraphics[width=0.9\textwidth]{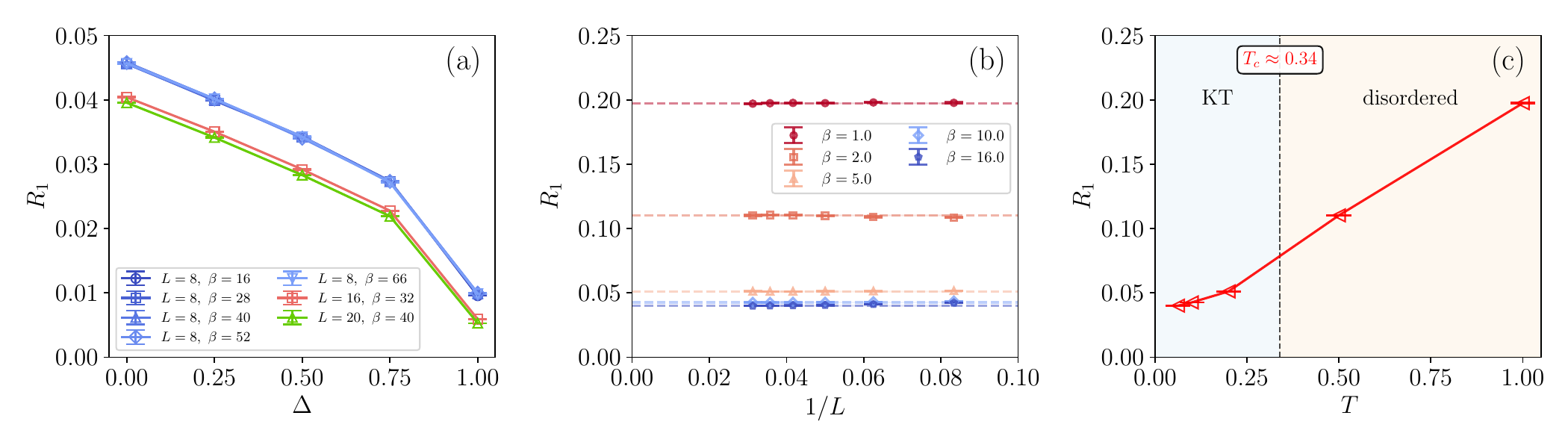}
\end{center}
\noindent {\small{FIG.~\thefigure.} R\'enyi-1 correlator $R_1(r=L/2)$ in the two-dimensional XXZ model.
(a) $R_1$ as a function of the anisotropy $\Delta$ for square lattices with $L=8, 16, 20$. In the ground-state limit ($\beta \gtrsim 2L$), $R_1$ maintains a finite value throughout the parameter space.
(b) Finite-size scaling of $R_1$ at the pure XY point ($\Delta=0$) across various temperatures. The correlator exhibits a distinct plateau in the large-$L$ regime.
(c) The dashed vertical line marks the Berezinskii-Kosterlitz-Thouless transition temperature $T_c \approx 0.34$. The signal evolves smoothly across $T_c$, and no behavioral changes are observed on either side of the phase transition. All data were computed using the GRDM framework.}
\label{fig:r1tot}
\vspace{3.0em}
\twocolumngrid

\textit{{\color{blue} Off-diagonal spectral function--}} As a direct application of the GRDM framework, we extract an off-diagonal spectral function by choosing a transverse insertion $S_i^x$, which is inaccessible in standard $S^z$-basis RDM sampling. We simulate the imaginary-time correlation function,
$C_{ij}(\tau)=\langle S_i^x(\tau)S_j^x(0)\rangle=\mathrm{Tr}\left(\tilde{\rho}_{i,j}^{S_i^x}(\tau)S_j^x\right)/\mathrm{Tr}\tilde{\rho}^{I}$ in 1D XXZ chain of $L=24$ at $\Delta=0.2$ with $\beta=4L$ to ensure ground-state convergence. The dynamical structure factor $S_{ij}(\omega)$ is extracted via stochastic analytic continuation (SAC)~\cite{Sandvik1998SAC, OlavFSAC, Sandvik2016SAC, HuiShaoSAC,yan2021topological}.  As shown in Fig.~\ref{fig:xxspectrum}, the low-energy spectral weight is concentrated around the antiferromagnetic momentum $k=\pi$, while the response displays a broadened continuum rather than a sharp quasiparticle branch. This behavior is consistent with the gapless critical regime of the spin-$1/2$ XXZ chain for $|\Delta|<1$, whose low-energy physics is described by a Luttinger liquid in the thermodynamic limit~\cite{Giamarchi2003}, and with the known continuum structure of the dynamical spin response in the anisotropic XXZ chain~\cite{Pereira2006}.

\textit{\color{blue}R\'enyi-1 correlator.---} The exploration of phase transitions in mixed states has recently extended the Landau paradigm by distinguishing between \textit{strong} and \textit{weak} symmetries~\cite{SWSSB_PRXQuantum, SWSSB_U1Hydrodynamics, SWSSBZack,SWSSB_Wightman,yuxuejiaSWSSB}. A mixed state $\rho = \sum_k p_k |\psi_k\rangle\langle\psi_k|$ possesses a strong symmetry if every pure-state component $|\psi_k\rangle$ carries the same symmetry charge, whereas a weak symmetry $[U, \rho] = 0$ only implies invariance of the ensemble average. A particularly subtle phenomenon is the strong-to-weak spontaneous symmetry breaking (SWSSB), where the strong symmetry is spontaneously lost while the weak symmetry remains intact. In this regime, conventional two-point correlation function decays to zero at long distances, thereby failing to detect the internal symmetry reorganization of the density matrix.

To diagnose SWSSB, non-linear measures such as the R\'enyi-1 (or Wightman) correlator $R_1$ are required~\cite{SWSSBZack,SWSSB_Wightman}:
\begin{equation}
    R_1(i,j) = \mathrm{Tr}\left[ \hat O_{ij} \sqrt{\rho} \hat O^\dagger_{ij} \sqrt{\rho} \right]
\end{equation}
 within a fixed symmetry sector, where $O_{ij} = O_i O_j^\dagger$ is defined by composite charged operator corresponding to the global symmetry of $\rho$. For thermal Gibbs states, the matrix square root in $R_1$ translates to an imaginary-time separation of $\beta/2$: 
$R_1(i,j) = \big\langle \hat O_{ij}(\tau=\beta/2) \hat O^\dagger_{ij}(\tau=0) \big\rangle_{\beta}$,
which can be directly implemented in the GRDM framework by inserting $\hat O_{ij}$ at $\tau=\beta/2$.

We study the two-dimensional spin-$1/2$ XXZ model, whose XY phase has a global $U(1)$ symmetry associated with total $S^z$ conservation. Although the Mermin-Wagner theorem forbids conventional spontaneous symmetry breaking at any finite temperature in two dimensions~\cite{MWT1966,Hohenberg1967,li2024relevant,wang2024sudden}, recent work proposed that a Gibbs state of a local symmetric Hamiltonian, when projected to a fixed charge sector, should nevertheless exhibit SWSSB in the absence of weak symmetry breaking~\cite{SWSSB_PRXQuantum}. To test this conjecture, we choose one off-diagonal observable $\hat{O}_{ij}=S_i^+S_j^-$ and $\hat{O}_{ij}^\dagger=S_i^-S_j^+$, and restrict the measurement to the sector with total $S^z=0$. Technical details and benchmark results are presented in Appendix~D.

We first examine the ground-state limit by scaling the inverse temperature as $\beta \ge 2L$. Under this condition, the R\'enyi-1 correlator converges, with larger values of $\beta$ yielding identical results. As shown in Fig.~\ref{fig:r1tot}, $R_1$ remains finite across the entire range of the anisotropy parameter $\Delta$. The magnitude of the signal is noticeably suppressed as the system approaches the isotropic Heisenberg point at $\Delta=1$.  \ Turning to the finite-temperature regime at the pure XY point ($\Delta=0$), we focus on the longest-distance correlator $R_1(1,L/2+1)$ and examine its finite-size analysis. The values of $R_1(L/2)$ exhibit a clear plateau as $L$ increases, indicating that the correlator converges to a non-zero constant in the thermodynamic limit. By fitting these convergent values, we further map the temperature dependence of $R_1$. Notably, the R\'enyi-1 signal remains smooth and shows no qualitative changes across the Berezinskii-Kosterlitz-Thouless transition point $T_c \approx 0.34$~\cite{BKT034}.  \ These results confirm the theoretical expectation that SWSSB order exists boardly at finite temperature. Consequently,  even when conventional correlators vanish at long distance, $R_1$ remains finite and thus quantifies the broken strong $U(1)$ symmetry encoded in the mixed-state density matrix.

\textit{\color{blue}Conclusion.---} We have introduced a GRDM-based framework that fundamentally extends the capability of QMC simulations. By integrating the boundary-hole trick with a systematic operator-insertion scheme, the method enables direct, polynomial-cost sampling of both equal-time and imaginary-time off-diagonal observables within a unified Monte Carlo scheme.

The framework is illustrated in two applications. First, we computed the transverse dynamical structure factor $S^{xx}(k,\omega)$, capturing gapless continua that are typically inaccessible in standard $S^z$-basis QMC. Second, we implemented an efficient estimator for the R\'enyi‑1 correlator $R_1(i,j)$, which diagnoses SWSSB in mixed states. Our large‑scale simulations of the two‑dimensional model provide direct numerical evidence that $R_1$ remains finite at finite temperature, confirming the robustness of SWSSB even in the absence of conventional long‑range order—a result that supports recent theoretical conjectures.

The GRDM formalism is not restricted to the two examples studied here. More broadly, it applies to any sign‑problem‑free QMC representation (SSE, path integral, etc.) and accommodates arbitrary local operators, whether diagonal or off‑diagonal. Beyond the observables presented here, the same framework grants access to quantities that were previously challenging to obtain. By providing a systematic way to ``open up'' the imaginary‑time manifold and insert operators at will, this work establishes a versatile paradigm for extracting universal information from quantum many‑body systems.

\textit{\color{blue} Acknowledgements.---} The authors thank Yi-Ming Ding, Yan-Cheng Wang, Chong Wang, and Meng Cheng for helpful discussions. 
This work is supported by the Scientific Research Project (No. WU2025B011) and the Start-up Funding of Westlake University. The authors thank the high-performance computing center of Westlake University for providing HPC resources.

\bibliography{refs}

\begin{thebibliography}{62}%
\makeatletter
\providecommand \@ifxundefined [1]{%
 \@ifx{#1\undefined}
}%
\providecommand \@ifnum [1]{%
 \ifnum #1\expandafter \@firstoftwo
 \else \expandafter \@secondoftwo
 \fi
}%
\providecommand \@ifx [1]{%
 \ifx #1\expandafter \@firstoftwo
 \else \expandafter \@secondoftwo
 \fi
}%
\providecommand \natexlab [1]{#1}%
\providecommand \enquote  [1]{``#1''}%
\providecommand \bibnamefont  [1]{#1}%
\providecommand \bibfnamefont [1]{#1}%
\providecommand \citenamefont [1]{#1}%
\providecommand \href@noop [0]{\@secondoftwo}%
\providecommand \href [0]{\begingroup \@sanitize@url \@href}%
\providecommand \@href[1]{\@@startlink{#1}\@@href}%
\providecommand \@@href[1]{\endgroup#1\@@endlink}%
\providecommand \@sanitize@url [0]{\catcode `\\12\catcode `\$12\catcode `\&12\catcode `\#12\catcode `\^12\catcode `\_12\catcode `\%12\relax}%
\providecommand \@@startlink[1]{}%
\providecommand \@@endlink[0]{}%
\providecommand \url  [0]{\begingroup\@sanitize@url \@url }%
\providecommand \@url [1]{\endgroup\@href {#1}{\urlprefix }}%
\providecommand \urlprefix  [0]{URL }%
\providecommand \Eprint [0]{\href }%
\providecommand \doibase [0]{https://doi.org/}%
\providecommand \selectlanguage [0]{\@gobble}%
\providecommand \bibinfo  [0]{\@secondoftwo}%
\providecommand \bibfield  [0]{\@secondoftwo}%
\providecommand \translation [1]{[#1]}%
\providecommand \BibitemOpen [0]{}%
\providecommand \bibitemStop [0]{}%
\providecommand \bibitemNoStop [0]{.\EOS\space}%
\providecommand \EOS [0]{\spacefactor3000\relax}%
\providecommand \BibitemShut  [1]{\csname bibitem#1\endcsname}%
\let\auto@bib@innerbib\@empty
\bibitem [{\citenamefont {Handscomb}(1962)}]{Handscomb_1962}%
  \BibitemOpen
  \bibfield  {author} {\bibinfo {author} {\bibfnamefont {D.~C.}\ \bibnamefont {Handscomb}},\ }\bibfield  {title} {\bibinfo {title} {The monte carlo method in quantum statistical mechanics},\ }\href {https://doi.org/10.1017/S0305004100040639} {\bibfield  {journal} {\bibinfo  {journal} {Mathematical Proceedings of the Cambridge Philosophical Society}\ }\textbf {\bibinfo {volume} {58}},\ \bibinfo {pages} {594} (\bibinfo {year} {1962})}\BibitemShut {NoStop}%
\bibitem [{\citenamefont {Blankenbecler}\ \emph {et~al.}(1981)\citenamefont {Blankenbecler}, \citenamefont {Scalapino},\ and\ \citenamefont {Sugar}}]{Blankenbecler1981Monte}%
  \BibitemOpen
  \bibfield  {author} {\bibinfo {author} {\bibfnamefont {R.}~\bibnamefont {Blankenbecler}}, \bibinfo {author} {\bibfnamefont {D.~J.}\ \bibnamefont {Scalapino}},\ and\ \bibinfo {author} {\bibfnamefont {R.~L.}\ \bibnamefont {Sugar}},\ }\bibfield  {title} {\bibinfo {title} {Monte carlo calculations of coupled boson-fermion systems. i},\ }\href {https://doi.org/10.1103/PhysRevD.24.2278} {\bibfield  {journal} {\bibinfo  {journal} {Phys. Rev. D}\ }\textbf {\bibinfo {volume} {24}},\ \bibinfo {pages} {2278} (\bibinfo {year} {1981})}\BibitemShut {NoStop}%
\bibitem [{\citenamefont {Scalapino}\ and\ \citenamefont {Sugar}(1981)}]{Scalapino1981Monte}%
  \BibitemOpen
  \bibfield  {author} {\bibinfo {author} {\bibfnamefont {D.~J.}\ \bibnamefont {Scalapino}}\ and\ \bibinfo {author} {\bibfnamefont {R.~L.}\ \bibnamefont {Sugar}},\ }\bibfield  {title} {\bibinfo {title} {Monte carlo calculations of coupled boson-fermion systems. ii},\ }\href {https://doi.org/10.1103/PhysRevB.24.4295} {\bibfield  {journal} {\bibinfo  {journal} {Phys. Rev. B}\ }\textbf {\bibinfo {volume} {24}},\ \bibinfo {pages} {4295} (\bibinfo {year} {1981})}\BibitemShut {NoStop}%
\bibitem [{\citenamefont {Hirsch}\ \emph {et~al.}(1982)\citenamefont {Hirsch}, \citenamefont {Sugar}, \citenamefont {Scalapino},\ and\ \citenamefont {Blankenbecler}}]{Hirsch1982}%
  \BibitemOpen
  \bibfield  {author} {\bibinfo {author} {\bibfnamefont {J.~E.}\ \bibnamefont {Hirsch}}, \bibinfo {author} {\bibfnamefont {R.~L.}\ \bibnamefont {Sugar}}, \bibinfo {author} {\bibfnamefont {D.~J.}\ \bibnamefont {Scalapino}},\ and\ \bibinfo {author} {\bibfnamefont {R.}~\bibnamefont {Blankenbecler}},\ }\bibfield  {title} {\bibinfo {title} {Monte carlo simulations of one-dimensional fermion systems},\ }\href {https://doi.org/10.1103/PhysRevB.26.5033} {\bibfield  {journal} {\bibinfo  {journal} {Phys. Rev. B}\ }\textbf {\bibinfo {volume} {26}},\ \bibinfo {pages} {5033} (\bibinfo {year} {1982})}\BibitemShut {NoStop}%
\bibitem [{\citenamefont {Hirsch}(1983)}]{Hirsch1983Discrete}%
  \BibitemOpen
  \bibfield  {author} {\bibinfo {author} {\bibfnamefont {J.~E.}\ \bibnamefont {Hirsch}},\ }\bibfield  {title} {\bibinfo {title} {Discrete hubbard-stratonovich transformation for fermion lattice models},\ }\href {https://doi.org/10.1103/PhysRevB.28.4059} {\bibfield  {journal} {\bibinfo  {journal} {Phys. Rev. B}\ }\textbf {\bibinfo {volume} {28}},\ \bibinfo {pages} {4059} (\bibinfo {year} {1983})}\BibitemShut {NoStop}%
\bibitem [{\citenamefont {Ceperley}\ and\ \citenamefont {Alder}(1986)}]{ceperley1986quantum}%
  \BibitemOpen
  \bibfield  {author} {\bibinfo {author} {\bibfnamefont {D.}~\bibnamefont {Ceperley}}\ and\ \bibinfo {author} {\bibfnamefont {B.}~\bibnamefont {Alder}},\ }\bibfield  {title} {\bibinfo {title} {Quantum monte carlo},\ }\href {https://doi.org/10.1126/science.231.4738.555} {\bibfield  {journal} {\bibinfo  {journal} {Science}\ }\textbf {\bibinfo {volume} {231}},\ \bibinfo {pages} {555} (\bibinfo {year} {1986})}\BibitemShut {NoStop}%
\bibitem [{\citenamefont {Sandvik}\ and\ \citenamefont {Kurkij\"arvi}(1991)}]{sandvik1991quantum}%
  \BibitemOpen
  \bibfield  {author} {\bibinfo {author} {\bibfnamefont {A.~W.}\ \bibnamefont {Sandvik}}\ and\ \bibinfo {author} {\bibfnamefont {J.}~\bibnamefont {Kurkij\"arvi}},\ }\bibfield  {title} {\bibinfo {title} {Quantum monte carlo simulation method for spin systems},\ }\href {https://doi.org/10.1103/PhysRevB.43.5950} {\bibfield  {journal} {\bibinfo  {journal} {Phys. Rev. B}\ }\textbf {\bibinfo {volume} {43}},\ \bibinfo {pages} {5950} (\bibinfo {year} {1991})}\BibitemShut {NoStop}%
\bibitem [{\citenamefont {Sandvik}(1999)}]{Sandvik1999}%
  \BibitemOpen
  \bibfield  {author} {\bibinfo {author} {\bibfnamefont {A.~W.}\ \bibnamefont {Sandvik}},\ }\bibfield  {title} {\bibinfo {title} {Stochastic series expansion method with operator-loop update},\ }\href {https://doi.org/10.1103/PhysRevB.59.R14157} {\bibfield  {journal} {\bibinfo  {journal} {Phys. Rev. B}\ }\textbf {\bibinfo {volume} {59}},\ \bibinfo {pages} {R14157} (\bibinfo {year} {1999})}\BibitemShut {NoStop}%
\bibitem [{\citenamefont {Foulkes}\ \emph {et~al.}(2001)\citenamefont {Foulkes}, \citenamefont {Mitas}, \citenamefont {Needs},\ and\ \citenamefont {Rajagopal}}]{Foulkes_RMP2001}%
  \BibitemOpen
  \bibfield  {author} {\bibinfo {author} {\bibfnamefont {W.~M.~C.}\ \bibnamefont {Foulkes}}, \bibinfo {author} {\bibfnamefont {L.}~\bibnamefont {Mitas}}, \bibinfo {author} {\bibfnamefont {R.~J.}\ \bibnamefont {Needs}},\ and\ \bibinfo {author} {\bibfnamefont {G.}~\bibnamefont {Rajagopal}},\ }\bibfield  {title} {\bibinfo {title} {Quantum monte carlo simulations of solids},\ }\href {https://doi.org/10.1103/RevModPhys.73.33} {\bibfield  {journal} {\bibinfo  {journal} {Rev. Mod. Phys.}\ }\textbf {\bibinfo {volume} {73}},\ \bibinfo {pages} {33} (\bibinfo {year} {2001})}\BibitemShut {NoStop}%
\bibitem [{\citenamefont {Sandvik}(2003{\natexlab{a}})}]{SandvikIsing}%
  \BibitemOpen
  \bibfield  {author} {\bibinfo {author} {\bibfnamefont {A.~W.}\ \bibnamefont {Sandvik}},\ }\bibfield  {title} {\bibinfo {title} {Stochastic series expansion method for quantum ising models with arbitrary interactions},\ }\href {https://doi.org/10.1103/PhysRevE.68.056701} {\bibfield  {journal} {\bibinfo  {journal} {Phys. Rev. E}\ }\textbf {\bibinfo {volume} {68}},\ \bibinfo {pages} {056701} (\bibinfo {year} {2003}{\natexlab{a}})}\BibitemShut {NoStop}%
\bibitem [{\citenamefont {Evertz}(2003)}]{Evertz2003loop}%
  \BibitemOpen
  \bibfield  {author} {\bibinfo {author} {\bibfnamefont {H.~G.}\ \bibnamefont {Evertz}},\ }\bibfield  {title} {\bibinfo {title} {The loop algorithm},\ }\href {https://doi.org/10.1080/0001873021000049195} {\bibfield  {journal} {\bibinfo  {journal} {Advances in Physics}\ }\textbf {\bibinfo {volume} {52}},\ \bibinfo {pages} {1} (\bibinfo {year} {2003})}\BibitemShut {NoStop}%
\bibitem [{\citenamefont {Assaad}\ and\ \citenamefont {Evertz}(2008)}]{Assaad_book2008}%
  \BibitemOpen
  \bibfield  {author} {\bibinfo {author} {\bibfnamefont {F.}~\bibnamefont {Assaad}}\ and\ \bibinfo {author} {\bibfnamefont {H.}~\bibnamefont {Evertz}},\ }\bibinfo {title} {World-line and determinantal quantum monte carlo methods for spins, phonons and electrons},\ in\ \href {https://doi.org/10.1007/978-3-540-74686-7_10} {\emph {\bibinfo {booktitle} {Computational Many-Particle Physics}}},\ \bibinfo {editor} {edited by\ \bibinfo {editor} {\bibfnamefont {H.}~\bibnamefont {Fehske}}, \bibinfo {editor} {\bibfnamefont {R.}~\bibnamefont {Schneider}},\ and\ \bibinfo {editor} {\bibfnamefont {A.}~\bibnamefont {Wei{\ss}e}}}\ (\bibinfo  {publisher} {Springer Berlin Heidelberg},\ \bibinfo {address} {Berlin, Heidelberg},\ \bibinfo {year} {2008})\ pp.\ \bibinfo {pages} {277--356}\BibitemShut {NoStop}%
\bibitem [{\citenamefont {Sandvik}(2010)}]{Sandvik2010Computational}%
  \BibitemOpen
  \bibfield  {author} {\bibinfo {author} {\bibfnamefont {A.~W.}\ \bibnamefont {Sandvik}},\ }\bibfield  {title} {\bibinfo {title} {Computational studies of quantum spin systems},\ }\href {https://doi.org/10.1063/1.3518900} {\bibfield  {journal} {\bibinfo  {journal} {AIP Conference Proceedings}\ }\textbf {\bibinfo {volume} {1297}},\ \bibinfo {pages} {135} (\bibinfo {year} {2010})}\BibitemShut {NoStop}%
\bibitem [{\citenamefont {Melko}(2013)}]{Melko2013Stochastic}%
  \BibitemOpen
  \bibfield  {author} {\bibinfo {author} {\bibfnamefont {R.~G.}\ \bibnamefont {Melko}},\ }\bibinfo {title} {Stochastic series expansion quantum monte carlo},\ in\ \href {https://doi.org/10.1007/978-3-642-35106-8_7} {\emph {\bibinfo {booktitle} {Strongly Correlated Systems: Numerical Methods}}},\ \bibinfo {editor} {edited by\ \bibinfo {editor} {\bibfnamefont {A.}~\bibnamefont {Avella}}\ and\ \bibinfo {editor} {\bibfnamefont {F.}~\bibnamefont {Mancini}}}\ (\bibinfo  {publisher} {Springer Berlin Heidelberg},\ \bibinfo {address} {Berlin, Heidelberg},\ \bibinfo {year} {2013})\ pp.\ \bibinfo {pages} {185--206}\BibitemShut {NoStop}%
\bibitem [{\citenamefont {Carlson}\ \emph {et~al.}(2015)\citenamefont {Carlson}, \citenamefont {Gandolfi}, \citenamefont {Pederiva}, \citenamefont {Pieper}, \citenamefont {Schiavilla}, \citenamefont {Schmidt},\ and\ \citenamefont {Wiringa}}]{Carlson_rmp2015}%
  \BibitemOpen
  \bibfield  {author} {\bibinfo {author} {\bibfnamefont {J.}~\bibnamefont {Carlson}}, \bibinfo {author} {\bibfnamefont {S.}~\bibnamefont {Gandolfi}}, \bibinfo {author} {\bibfnamefont {F.}~\bibnamefont {Pederiva}}, \bibinfo {author} {\bibfnamefont {S.~C.}\ \bibnamefont {Pieper}}, \bibinfo {author} {\bibfnamefont {R.}~\bibnamefont {Schiavilla}}, \bibinfo {author} {\bibfnamefont {K.~E.}\ \bibnamefont {Schmidt}},\ and\ \bibinfo {author} {\bibfnamefont {R.~B.}\ \bibnamefont {Wiringa}},\ }\bibfield  {title} {\bibinfo {title} {Quantum monte carlo methods for nuclear physics},\ }\href {https://doi.org/10.1103/RevModPhys.87.1067} {\bibfield  {journal} {\bibinfo  {journal} {Rev. Mod. Phys.}\ }\textbf {\bibinfo {volume} {87}},\ \bibinfo {pages} {1067} (\bibinfo {year} {2015})}\BibitemShut {NoStop}%
\bibitem [{\citenamefont {Gubernatis}\ \emph {et~al.}(2016)\citenamefont {Gubernatis}, \citenamefont {Kawashima},\ and\ \citenamefont {Werner}}]{gubernatis2016quantum}%
  \BibitemOpen
  \bibfield  {author} {\bibinfo {author} {\bibfnamefont {J.}~\bibnamefont {Gubernatis}}, \bibinfo {author} {\bibfnamefont {N.}~\bibnamefont {Kawashima}},\ and\ \bibinfo {author} {\bibfnamefont {P.}~\bibnamefont {Werner}},\ }\href {https://doi.org/10.1017/CBO9780511902581} {\emph {\bibinfo {title} {Quantum Monte Carlo Methods}}}\ (\bibinfo  {publisher} {Cambridge University Press},\ \bibinfo {year} {2016})\BibitemShut {NoStop}%
\bibitem [{\citenamefont {Yan}\ \emph {et~al.}(2019)\citenamefont {Yan}, \citenamefont {Wu}, \citenamefont {Liu}, \citenamefont {Sylju\aa{}sen}, \citenamefont {Lou},\ and\ \citenamefont {Chen}}]{Yan2019}%
  \BibitemOpen
  \bibfield  {author} {\bibinfo {author} {\bibfnamefont {Z.}~\bibnamefont {Yan}}, \bibinfo {author} {\bibfnamefont {Y.}~\bibnamefont {Wu}}, \bibinfo {author} {\bibfnamefont {C.}~\bibnamefont {Liu}}, \bibinfo {author} {\bibfnamefont {O.~F.}\ \bibnamefont {Sylju\aa{}sen}}, \bibinfo {author} {\bibfnamefont {J.}~\bibnamefont {Lou}},\ and\ \bibinfo {author} {\bibfnamefont {Y.}~\bibnamefont {Chen}},\ }\bibfield  {title} {\bibinfo {title} {Sweeping cluster algorithm for quantum spin systems with strong geometric restrictions},\ }\href {https://doi.org/10.1103/PhysRevB.99.165135} {\bibfield  {journal} {\bibinfo  {journal} {Phys. Rev. B}\ }\textbf {\bibinfo {volume} {99}},\ \bibinfo {pages} {165135} (\bibinfo {year} {2019})}\BibitemShut {NoStop}%
\bibitem [{\citenamefont {Yan}(2022)}]{ZY2020improved}%
  \BibitemOpen
  \bibfield  {author} {\bibinfo {author} {\bibfnamefont {Z.}~\bibnamefont {Yan}},\ }\bibfield  {title} {\bibinfo {title} {Global scheme of sweeping cluster algorithm to sample among topological sectors},\ }\href {https://doi.org/10.1103/PhysRevB.105.184432} {\bibfield  {journal} {\bibinfo  {journal} {Phys. Rev. B}\ }\textbf {\bibinfo {volume} {105}},\ \bibinfo {pages} {184432} (\bibinfo {year} {2022})}\BibitemShut {NoStop}%
\bibitem [{\citenamefont {Sandvik}(1992)}]{sandvik1992generalization}%
  \BibitemOpen
  \bibfield  {author} {\bibinfo {author} {\bibfnamefont {A.~W.}\ \bibnamefont {Sandvik}},\ }\bibfield  {title} {\bibinfo {title} {A generalization of handscomb's quantum monte carlo scheme-application to the 1d hubbard model},\ }\href {https://doi.org/10.1088/0305-4470/25/13/017} {\bibfield  {journal} {\bibinfo  {journal} {Journal of Physics A: Mathematical and General}\ }\textbf {\bibinfo {volume} {25}},\ \bibinfo {pages} {3667} (\bibinfo {year} {1992})}\BibitemShut {NoStop}%
\bibitem [{\citenamefont {Yan}\ \emph {et~al.}(2023)\citenamefont {Yan}, \citenamefont {Wang}, \citenamefont {Samajdar}, \citenamefont {Sachdev},\ and\ \citenamefont {Meng}}]{yan2023emergent}%
  \BibitemOpen
  \bibfield  {author} {\bibinfo {author} {\bibfnamefont {Z.}~\bibnamefont {Yan}}, \bibinfo {author} {\bibfnamefont {Y.-C.}\ \bibnamefont {Wang}}, \bibinfo {author} {\bibfnamefont {R.}~\bibnamefont {Samajdar}}, \bibinfo {author} {\bibfnamefont {S.}~\bibnamefont {Sachdev}},\ and\ \bibinfo {author} {\bibfnamefont {Z.~Y.}\ \bibnamefont {Meng}},\ }\bibfield  {title} {\bibinfo {title} {Emergent glassy behavior in a kagome rydberg atom array},\ }\href {https://doi.org/10.1103/PhysRevLett.130.206501} {\bibfield  {journal} {\bibinfo  {journal} {Phys. Rev. Lett.}\ }\textbf {\bibinfo {volume} {130}},\ \bibinfo {pages} {206501} (\bibinfo {year} {2023})}\BibitemShut {NoStop}%
\bibitem [{\citenamefont {{Yan}}\ \emph {et~al.}(2022)\citenamefont {{Yan}}, \citenamefont {{Samajdar}}, \citenamefont {{Wang}}, \citenamefont {{Sachdev}},\ and\ \citenamefont {{Meng}}}]{ZYan2022}%
  \BibitemOpen
  \bibfield  {author} {\bibinfo {author} {\bibfnamefont {Z.}~\bibnamefont {{Yan}}}, \bibinfo {author} {\bibfnamefont {R.}~\bibnamefont {{Samajdar}}}, \bibinfo {author} {\bibfnamefont {Y.-C.}\ \bibnamefont {{Wang}}}, \bibinfo {author} {\bibfnamefont {S.}~\bibnamefont {{Sachdev}}},\ and\ \bibinfo {author} {\bibfnamefont {Z.~Y.}\ \bibnamefont {{Meng}}},\ }\bibfield  {title} {\bibinfo {title} {{Triangular lattice quantum dimer model with variable dimer density}},\ }\href {https://www.nature.com/articles/s41467-022-33431-5} {\bibfield  {journal} {\bibinfo  {journal} {Nat. Commun.}\ }\textbf {\bibinfo {volume} {13}},\ \bibinfo {pages} {5799} (\bibinfo {year} {2022})}\BibitemShut {NoStop}%
\bibitem [{\citenamefont {Prokof'ev}\ \emph {et~al.}(1998)\citenamefont {Prokof'ev}, \citenamefont {Svistunov},\ and\ \citenamefont {Tupitsyn}}]{NProkofev1998}%
  \BibitemOpen
  \bibfield  {author} {\bibinfo {author} {\bibfnamefont {N.~V.}\ \bibnamefont {Prokof'ev}}, \bibinfo {author} {\bibfnamefont {B.~V.}\ \bibnamefont {Svistunov}},\ and\ \bibinfo {author} {\bibfnamefont {I.~S.}\ \bibnamefont {Tupitsyn}},\ }\bibfield  {title} {\bibinfo {title} {{Exact, complete, and universal continuous-time worldline Monte Carlo approach to the statistics of discrete quantum systems}},\ }\href {https://doi.org/10.1134/1.558661} {\bibfield  {journal} {\bibinfo  {journal} {J. Exp. Theor. Phys.}\ }\textbf {\bibinfo {volume} {87}},\ \bibinfo {pages} {310} (\bibinfo {year} {1998})},\ \Eprint {https://arxiv.org/abs/cond-mat/9703200} {arXiv:cond-mat/9703200} \BibitemShut {NoStop}%
\bibitem [{\citenamefont {Sylju\aa{}sen}\ and\ \citenamefont {Sandvik}(2002)}]{DL1_OlavF_2002}%
  \BibitemOpen
  \bibfield  {author} {\bibinfo {author} {\bibfnamefont {O.~F.}\ \bibnamefont {Sylju\aa{}sen}}\ and\ \bibinfo {author} {\bibfnamefont {A.~W.}\ \bibnamefont {Sandvik}},\ }\bibfield  {title} {\bibinfo {title} {Quantum monte carlo with directed loops},\ }\href {https://doi.org/10.1103/PhysRevE.66.046701} {\bibfield  {journal} {\bibinfo  {journal} {Phys. Rev. E}\ }\textbf {\bibinfo {volume} {66}},\ \bibinfo {pages} {046701} (\bibinfo {year} {2002})}\BibitemShut {NoStop}%
\bibitem [{\citenamefont {Dorneich}\ and\ \citenamefont {Troyer}(2001)}]{Dorneich2001}%
  \BibitemOpen
  \bibfield  {author} {\bibinfo {author} {\bibfnamefont {A.}~\bibnamefont {Dorneich}}\ and\ \bibinfo {author} {\bibfnamefont {M.}~\bibnamefont {Troyer}},\ }\bibfield  {title} {\bibinfo {title} {Accessing the dynamics of large many-particle systems using the stochastic series expansion},\ }\href {https://doi.org/10.1103/PhysRevE.64.066701} {\bibfield  {journal} {\bibinfo  {journal} {Phys. Rev. E}\ }\textbf {\bibinfo {volume} {64}},\ \bibinfo {pages} {066701} (\bibinfo {year} {2001})}\BibitemShut {NoStop}%
\bibitem [{\citenamefont {Alet}\ \emph {et~al.}(2005)\citenamefont {Alet}, \citenamefont {Wessel},\ and\ \citenamefont {Troyer}}]{alet2005generalized}%
  \BibitemOpen
  \bibfield  {author} {\bibinfo {author} {\bibfnamefont {F.}~\bibnamefont {Alet}}, \bibinfo {author} {\bibfnamefont {S.}~\bibnamefont {Wessel}},\ and\ \bibinfo {author} {\bibfnamefont {M.}~\bibnamefont {Troyer}},\ }\bibfield  {title} {\bibinfo {title} {Generalized directed loop method for quantum monte carlo simulations},\ }\href {https://doi.org/10.1103/PhysRevE.71.036706} {\bibfield  {journal} {\bibinfo  {journal} {Phys. Rev. E}\ }\textbf {\bibinfo {volume} {71}},\ \bibinfo {pages} {036706} (\bibinfo {year} {2005})}\BibitemShut {NoStop}%
\bibitem [{\citenamefont {Kashurnikov}\ \emph {et~al.}(1999)\citenamefont {Kashurnikov}, \citenamefont {Prokof'ev}, \citenamefont {Svistunov},\ and\ \citenamefont {Troyer}}]{NProkofev1999}%
  \BibitemOpen
  \bibfield  {author} {\bibinfo {author} {\bibfnamefont {V.~A.}\ \bibnamefont {Kashurnikov}}, \bibinfo {author} {\bibfnamefont {N.~V.}\ \bibnamefont {Prokof'ev}}, \bibinfo {author} {\bibfnamefont {B.~V.}\ \bibnamefont {Svistunov}},\ and\ \bibinfo {author} {\bibfnamefont {M.}~\bibnamefont {Troyer}},\ }\bibfield  {title} {\bibinfo {title} {Quantum spin chains in a magnetic field},\ }\href {https://doi.org/10.1103/PhysRevB.59.1162} {\bibfield  {journal} {\bibinfo  {journal} {Phys. Rev. B}\ }\textbf {\bibinfo {volume} {59}},\ \bibinfo {pages} {1162} (\bibinfo {year} {1999})}\BibitemShut {NoStop}%
\bibitem [{\citenamefont {Sylju\aa{}sen}(2003)}]{DL2_OlavF_2003}%
  \BibitemOpen
  \bibfield  {author} {\bibinfo {author} {\bibfnamefont {O.~F.}\ \bibnamefont {Sylju\aa{}sen}},\ }\bibfield  {title} {\bibinfo {title} {Directed loop updates for quantum lattice models},\ }\href {https://doi.org/10.1103/PhysRevE.67.046701} {\bibfield  {journal} {\bibinfo  {journal} {Phys. Rev. E}\ }\textbf {\bibinfo {volume} {67}},\ \bibinfo {pages} {046701} (\bibinfo {year} {2003})}\BibitemShut {NoStop}%
\bibitem [{\citenamefont {Zhu}\ and\ \citenamefont {Guo}(2021)}]{wenjing2021measuring}%
  \BibitemOpen
  \bibfield  {author} {\bibinfo {author} {\bibfnamefont {W.}~\bibnamefont {Zhu}}\ and\ \bibinfo {author} {\bibfnamefont {W.}~\bibnamefont {Guo}},\ }\bibfield  {title} {\bibinfo {title} {Measuring off-diagonal correlation function in stochastic series expansion quantum monte carlo simulation},\ }\href {https://doi.org/10.12202/j.0476-0301.2021061} {\bibfield  {journal} {\bibinfo  {journal} {Journal of Beijing Normal University (Natural Science)}\ }\textbf {\bibinfo {volume} {57}},\ \bibinfo {pages} {593} (\bibinfo {year} {2021})}\BibitemShut {NoStop}%
\bibitem [{\citenamefont {Zhou}\ \emph {et~al.}(2025)\citenamefont {Zhou}, \citenamefont {Zhou}, \citenamefont {Desrochers}, \citenamefont {Kim},\ and\ \citenamefont {Meng}}]{multiDL2025}%
  \BibitemOpen
  \bibfield  {author} {\bibinfo {author} {\bibfnamefont {C.}~\bibnamefont {Zhou}}, \bibinfo {author} {\bibfnamefont {Z.}~\bibnamefont {Zhou}}, \bibinfo {author} {\bibfnamefont {F.}~\bibnamefont {Desrochers}}, \bibinfo {author} {\bibfnamefont {Y.~B.}\ \bibnamefont {Kim}},\ and\ \bibinfo {author} {\bibfnamefont {Z.~Y.}\ \bibnamefont {Meng}},\ }\href {https://arxiv.org/abs/2510.14813} {\bibinfo {title} {Quantum fisher information as a thermal and dynamical probe in frustrated magnets: Insights from quantum spin ice}} (\bibinfo {year} {2025}),\ \Eprint {https://arxiv.org/abs/2510.14813} {arXiv:2510.14813 [cond-mat.str-el]} \BibitemShut {NoStop}%
\bibitem [{\citenamefont {Tarabunga}\ and\ \citenamefont {Ding}(2025)}]{ymding2025bell}%
  \BibitemOpen
  \bibfield  {author} {\bibinfo {author} {\bibfnamefont {P.~S.}\ \bibnamefont {Tarabunga}}\ and\ \bibinfo {author} {\bibfnamefont {Y.-M.}\ \bibnamefont {Ding}},\ }\bibfield  {title} {\bibinfo {title} {Bell sampling in quantum monte carlo simulations},\ }\href {https://doi.org/10.1103/fq8z-y55j} {\bibfield  {journal} {\bibinfo  {journal} {Phys. Rev. Lett.}\ }\textbf {\bibinfo {volume} {135}},\ \bibinfo {pages} {200403} (\bibinfo {year} {2025})}\BibitemShut {NoStop}%
\bibitem [{\citenamefont {Wang}\ \emph {et~al.}(2026)\citenamefont {Wang}, \citenamefont {Liu}, \citenamefont {Mao}, \citenamefont {Wang},\ and\ \citenamefont {Yan}}]{BRAZhiyan}%
  \BibitemOpen
  \bibfield  {author} {\bibinfo {author} {\bibfnamefont {Z.}~\bibnamefont {Wang}}, \bibinfo {author} {\bibfnamefont {Z.}~\bibnamefont {Liu}}, \bibinfo {author} {\bibfnamefont {B.-B.}\ \bibnamefont {Mao}}, \bibinfo {author} {\bibfnamefont {Z.}~\bibnamefont {Wang}},\ and\ \bibinfo {author} {\bibfnamefont {Z.}~\bibnamefont {Yan}},\ }\bibfield  {title} {\bibinfo {title} {Addressing general measurements in quantum monte carlo},\ }\href {https://doi.org/10.1038/s41467-025-67324-0} {\bibfield  {journal} {\bibinfo  {journal} {Nature Communications}\ }\textbf {\bibinfo {volume} {17}},\ \bibinfo {pages} {679} (\bibinfo {year} {2026})}\BibitemShut {NoStop}%
\bibitem [{\citenamefont {Mao}\ \emph {et~al.}(2025)\citenamefont {Mao}, \citenamefont {Ding}, \citenamefont {Wang}, \citenamefont {Hu},\ and\ \citenamefont {Yan}}]{Mao2025}%
  \BibitemOpen
  \bibfield  {author} {\bibinfo {author} {\bibfnamefont {B.-B.}\ \bibnamefont {Mao}}, \bibinfo {author} {\bibfnamefont {Y.-M.}\ \bibnamefont {Ding}}, \bibinfo {author} {\bibfnamefont {Z.}~\bibnamefont {Wang}}, \bibinfo {author} {\bibfnamefont {S.}~\bibnamefont {Hu}},\ and\ \bibinfo {author} {\bibfnamefont {Z.}~\bibnamefont {Yan}},\ }\bibfield  {title} {\bibinfo {title} {Sampling reduced density matrix to extract fine levels of entanglement spectrum and restore entanglement hamiltonian},\ }\href {https://doi.org/10.1038/s41467-025-58058-0} {\bibfield  {journal} {\bibinfo  {journal} {Nature Communications}\ }\textbf {\bibinfo {volume} {16}},\ \bibinfo {pages} {2880} (\bibinfo {year} {2025})}\BibitemShut {NoStop}%
\bibitem [{\citenamefont {Mao}\ \emph {et~al.}(2026)\citenamefont {Mao}, \citenamefont {Wang}, \citenamefont {Chen},\ and\ \citenamefont {Yan}}]{mao2025detecting}%
  \BibitemOpen
  \bibfield  {author} {\bibinfo {author} {\bibfnamefont {B.-B.}\ \bibnamefont {Mao}}, \bibinfo {author} {\bibfnamefont {Z.}~\bibnamefont {Wang}}, \bibinfo {author} {\bibfnamefont {B.-B.}\ \bibnamefont {Chen}},\ and\ \bibinfo {author} {\bibfnamefont {Z.}~\bibnamefont {Yan}},\ }\bibfield  {title} {\bibinfo {title} {Detecting the emergent continuous symmetry of criticality via a subsystem's entanglement spectrum},\ }\href {https://doi.org/10.1103/7j21-l3pg} {\bibfield  {journal} {\bibinfo  {journal} {Phys. Rev. Lett.}\ }\textbf {\bibinfo {volume} {136}},\ \bibinfo {pages} {046401} (\bibinfo {year} {2026})}\BibitemShut {NoStop}%
\bibitem [{\citenamefont {Wang}\ \emph {et~al.}(2025{\natexlab{a}})\citenamefont {Wang}, \citenamefont {Song}, \citenamefont {Lyu}, \citenamefont {Witczak-Krempa},\ and\ \citenamefont {Meng}}]{wang2025entanglement}%
  \BibitemOpen
  \bibfield  {author} {\bibinfo {author} {\bibfnamefont {T.-T.}\ \bibnamefont {Wang}}, \bibinfo {author} {\bibfnamefont {M.}~\bibnamefont {Song}}, \bibinfo {author} {\bibfnamefont {L.}~\bibnamefont {Lyu}}, \bibinfo {author} {\bibfnamefont {W.}~\bibnamefont {Witczak-Krempa}},\ and\ \bibinfo {author} {\bibfnamefont {Z.~Y.}\ \bibnamefont {Meng}},\ }\bibfield  {title} {\bibinfo {title} {Entanglement microscopy and tomography in many-body systems},\ }\href {https://doi.org/10.1038/s41467-024-55354-z} {\bibfield  {journal} {\bibinfo  {journal} {Nature Communications}\ }\textbf {\bibinfo {volume} {16}},\ \bibinfo {pages} {96} (\bibinfo {year} {2025}{\natexlab{a}})}\BibitemShut {NoStop}%
\bibitem [{\citenamefont {Chincholi}\ \emph {et~al.}(2026)\citenamefont {Chincholi}, \citenamefont {Capponi},\ and\ \citenamefont {Alet}}]{Fabien2025CDM}%
  \BibitemOpen
  \bibfield  {author} {\bibinfo {author} {\bibfnamefont {A.}~\bibnamefont {Chincholi}}, \bibinfo {author} {\bibfnamefont {S.}~\bibnamefont {Capponi}},\ and\ \bibinfo {author} {\bibfnamefont {F.}~\bibnamefont {Alet}},\ }\bibfield  {title} {\bibinfo {title} {Detecting the largest correlations for many-body systems using the correlation density matrix: A quantum monte carlo approach},\ }\href {https://doi.org/10.1103/fn6m-fbw2} {\bibfield  {journal} {\bibinfo  {journal} {Phys. Rev. B}\ }\textbf {\bibinfo {volume} {113}},\ \bibinfo {pages} {035122} (\bibinfo {year} {2026})}\BibitemShut {NoStop}%
\bibitem [{\citenamefont {Song}\ \emph {et~al.}(2025)\citenamefont {Song}, \citenamefont {Wang}, \citenamefont {Lyu}, \citenamefont {Witczak-Krempa},\ and\ \citenamefont {Meng}}]{Mengziyang2025b}%
  \BibitemOpen
  \bibfield  {author} {\bibinfo {author} {\bibfnamefont {M.}~\bibnamefont {Song}}, \bibinfo {author} {\bibfnamefont {T.-T.}\ \bibnamefont {Wang}}, \bibinfo {author} {\bibfnamefont {L.}~\bibnamefont {Lyu}}, \bibinfo {author} {\bibfnamefont {W.}~\bibnamefont {Witczak-Krempa}},\ and\ \bibinfo {author} {\bibfnamefont {Z.~Y.}\ \bibnamefont {Meng}},\ }\href {https://arxiv.org/abs/2509.09983} {\bibinfo {title} {Entanglement architecture of beyond-landau quantum criticality}} (\bibinfo {year} {2025}),\ \Eprint {https://arxiv.org/abs/2509.09983} {arXiv:2509.09983 [cond-mat.str-el]} \BibitemShut {NoStop}%
\bibitem [{\citenamefont {Jiang}\ \emph {et~al.}(2025)\citenamefont {Jiang}, \citenamefont {Luo}, \citenamefont {Mao},\ and\ \citenamefont {Yan}}]{Jiang2026indentifying}%
  \BibitemOpen
  \bibfield  {author} {\bibinfo {author} {\bibfnamefont {W.}~\bibnamefont {Jiang}}, \bibinfo {author} {\bibfnamefont {X.}~\bibnamefont {Luo}}, \bibinfo {author} {\bibfnamefont {B.-B.}\ \bibnamefont {Mao}},\ and\ \bibinfo {author} {\bibfnamefont {Z.}~\bibnamefont {Yan}},\ }\bibfield  {title} {\bibinfo {title} {Identifying the two-dimensional topological phase transition by entanglement spectrum: a fermion monte carlo study},\ }\href {https://doi.org/10.1088/1572-9494/ae1e65} {\bibfield  {journal} {\bibinfo  {journal} {Communications in Theoretical Physics}\ }\textbf {\bibinfo {volume} {78}},\ \bibinfo {pages} {045701} (\bibinfo {year} {2025})}\BibitemShut {NoStop}%
\bibitem [{\citenamefont {Timsina}\ \emph {et~al.}(2025)\citenamefont {Timsina}, \citenamefont {Ding}, \citenamefont {Tirrito}, \citenamefont {Tarabunga}, \citenamefont {Mao}, \citenamefont {Collura}, \citenamefont {Yan},\ and\ \citenamefont {Dalmonte}}]{Timsina2025robustness}%
  \BibitemOpen
  \bibfield  {author} {\bibinfo {author} {\bibfnamefont {H.}~\bibnamefont {Timsina}}, \bibinfo {author} {\bibfnamefont {Y.-M.}\ \bibnamefont {Ding}}, \bibinfo {author} {\bibfnamefont {E.}~\bibnamefont {Tirrito}}, \bibinfo {author} {\bibfnamefont {P.~S.}\ \bibnamefont {Tarabunga}}, \bibinfo {author} {\bibfnamefont {B.-B.}\ \bibnamefont {Mao}}, \bibinfo {author} {\bibfnamefont {M.}~\bibnamefont {Collura}}, \bibinfo {author} {\bibfnamefont {Z.}~\bibnamefont {Yan}},\ and\ \bibinfo {author} {\bibfnamefont {M.}~\bibnamefont {Dalmonte}},\ }\bibfield  {title} {\bibinfo {title} {Robustness of nonstabilizerness in the quantum ising chain via quantum monte carlo tomography},\ }\href {https://doi.org/10.1103/4hpw-6mq3} {\bibfield  {journal} {\bibinfo  {journal} {Phys. Rev. B}\ }\textbf {\bibinfo {volume} {112}},\ \bibinfo {pages} {165135} (\bibinfo {year} {2025})}\BibitemShut {NoStop}%
\bibitem [{Note1()}]{Note1}%
  \BibitemOpen
  \bibinfo {note} {Because all the systems studied so far are based cluster or operator-loop scheme of SSE with fixed update-configurations due to a good symmetry, such as TFIMs and Heisenberg models, while the update scheme is uncertain with a probability-based update-path (e.g., XXZ models), the previous methods lose effectiveness.}\BibitemShut {Stop}%
\bibitem [{\citenamefont {Weinstein}(2025)}]{SWSSBZack}%
  \BibitemOpen
  \bibfield  {author} {\bibinfo {author} {\bibfnamefont {Z.}~\bibnamefont {Weinstein}},\ }\bibfield  {title} {\bibinfo {title} {Efficient detection of strong-to-weak spontaneous symmetry breaking via the r\'enyi-1 correlator},\ }\href {https://doi.org/10.1103/PhysRevLett.134.150405} {\bibfield  {journal} {\bibinfo  {journal} {Phys. Rev. Lett.}\ }\textbf {\bibinfo {volume} {134}},\ \bibinfo {pages} {150405} (\bibinfo {year} {2025})}\BibitemShut {NoStop}%
\bibitem [{\citenamefont {Liu}\ \emph {et~al.}(2025)\citenamefont {Liu}, \citenamefont {Chen}, \citenamefont {Zhang}, \citenamefont {Zhou},\ and\ \citenamefont {Zhang}}]{SWSSB_Wightman}%
  \BibitemOpen
  \bibfield  {author} {\bibinfo {author} {\bibfnamefont {Z.}~\bibnamefont {Liu}}, \bibinfo {author} {\bibfnamefont {L.}~\bibnamefont {Chen}}, \bibinfo {author} {\bibfnamefont {Y.}~\bibnamefont {Zhang}}, \bibinfo {author} {\bibfnamefont {S.}~\bibnamefont {Zhou}},\ and\ \bibinfo {author} {\bibfnamefont {P.}~\bibnamefont {Zhang}},\ }\bibfield  {title} {\bibinfo {title} {Diagnosing strong-to-weak symmetry breaking via wightman correlators},\ }\href {https://doi.org/10.1038/s42005-025-02199-7} {\bibfield  {journal} {\bibinfo  {journal} {Communications Physics}\ }\textbf {\bibinfo {volume} {8}},\ \bibinfo {pages} {274} (\bibinfo {year} {2025})}\BibitemShut {NoStop}%
\bibitem [{\citenamefont {Lessa}\ \emph {et~al.}(2025)\citenamefont {Lessa}, \citenamefont {Ma}, \citenamefont {Zhang}, \citenamefont {Bi}, \citenamefont {Cheng},\ and\ \citenamefont {Wang}}]{SWSSB_PRXQuantum}%
  \BibitemOpen
  \bibfield  {author} {\bibinfo {author} {\bibfnamefont {L.~A.}\ \bibnamefont {Lessa}}, \bibinfo {author} {\bibfnamefont {R.}~\bibnamefont {Ma}}, \bibinfo {author} {\bibfnamefont {J.-H.}\ \bibnamefont {Zhang}}, \bibinfo {author} {\bibfnamefont {Z.}~\bibnamefont {Bi}}, \bibinfo {author} {\bibfnamefont {M.}~\bibnamefont {Cheng}},\ and\ \bibinfo {author} {\bibfnamefont {C.}~\bibnamefont {Wang}},\ }\bibfield  {title} {\bibinfo {title} {Strong-to-weak spontaneous symmetry breaking in mixed quantum states},\ }\href {https://doi.org/10.1103/PRXQuantum.6.010344} {\bibfield  {journal} {\bibinfo  {journal} {PRX Quantum}\ }\textbf {\bibinfo {volume} {6}},\ \bibinfo {pages} {010344} (\bibinfo {year} {2025})}\BibitemShut {NoStop}%
\bibitem [{\citenamefont {Troyer}\ \emph {et~al.}(2003)\citenamefont {Troyer}, \citenamefont {Alet}, \citenamefont {Trebst},\ and\ \citenamefont {Wessel}}]{Troyer2003nonlocalupdate}%
  \BibitemOpen
  \bibfield  {author} {\bibinfo {author} {\bibfnamefont {M.}~\bibnamefont {Troyer}}, \bibinfo {author} {\bibfnamefont {F.}~\bibnamefont {Alet}}, \bibinfo {author} {\bibfnamefont {S.}~\bibnamefont {Trebst}},\ and\ \bibinfo {author} {\bibfnamefont {S.}~\bibnamefont {Wessel}},\ }\bibfield  {title} {\bibinfo {title} {Non‐local updates for quantum monte carlo simulations},\ }\href {https://doi.org/10.1063/1.1632126} {\bibfield  {journal} {\bibinfo  {journal} {AIP Conference Proceedings}\ }\textbf {\bibinfo {volume} {690}},\ \bibinfo {pages} {156} (\bibinfo {year} {2003})}\BibitemShut {NoStop}%
\bibitem [{\citenamefont {Bl\"ote}\ and\ \citenamefont {Deng}(2002)}]{Deng2002Cluster}%
  \BibitemOpen
  \bibfield  {author} {\bibinfo {author} {\bibfnamefont {H.~W.~J.}\ \bibnamefont {Bl\"ote}}\ and\ \bibinfo {author} {\bibfnamefont {Y.}~\bibnamefont {Deng}},\ }\bibfield  {title} {\bibinfo {title} {Cluster monte carlo simulation of the transverse ising model},\ }\href {https://doi.org/10.1103/PhysRevE.66.066110} {\bibfield  {journal} {\bibinfo  {journal} {Phys. Rev. E}\ }\textbf {\bibinfo {volume} {66}},\ \bibinfo {pages} {066110} (\bibinfo {year} {2002})}\BibitemShut {NoStop}%
\bibitem [{\citenamefont {Huang}\ \emph {et~al.}(2020)\citenamefont {Huang}, \citenamefont {Liu}, \citenamefont {Jiang},\ and\ \citenamefont {Deng}}]{Huang2020Worm}%
  \BibitemOpen
  \bibfield  {author} {\bibinfo {author} {\bibfnamefont {C.-J.}\ \bibnamefont {Huang}}, \bibinfo {author} {\bibfnamefont {L.}~\bibnamefont {Liu}}, \bibinfo {author} {\bibfnamefont {Y.}~\bibnamefont {Jiang}},\ and\ \bibinfo {author} {\bibfnamefont {Y.}~\bibnamefont {Deng}},\ }\bibfield  {title} {\bibinfo {title} {Worm-algorithm-type simulation of the quantum transverse-field ising model},\ }\href {https://doi.org/10.1103/PhysRevB.102.094101} {\bibfield  {journal} {\bibinfo  {journal} {Phys. Rev. B}\ }\textbf {\bibinfo {volume} {102}},\ \bibinfo {pages} {094101} (\bibinfo {year} {2020})}\BibitemShut {NoStop}%
\bibitem [{\citenamefont {Sandvik}(2003{\natexlab{b}})}]{Sandvik2003Stochastic}%
  \BibitemOpen
  \bibfield  {author} {\bibinfo {author} {\bibfnamefont {A.~W.}\ \bibnamefont {Sandvik}},\ }\bibfield  {title} {\bibinfo {title} {Stochastic series expansion method for quantum ising models with arbitrary interactions},\ }\href {https://doi.org/10.1103/PhysRevE.68.056701} {\bibfield  {journal} {\bibinfo  {journal} {Phys. Rev. E}\ }\textbf {\bibinfo {volume} {68}},\ \bibinfo {pages} {056701} (\bibinfo {year} {2003}{\natexlab{b}})}\BibitemShut {NoStop}%
\bibitem [{Note2()}]{Note2}%
  \BibitemOpen
  \bibinfo {note} {Here the horizontal axis is still labeled by $\tau $ to indicate that each $\protect \tilde {\rho }_A^{I}$ is obtained together with the corresponding operator insertion at imaginary time $\tau $. As expected, the resulting equal-time correlation functions show that the sampled $\protect \tilde {\rho }_A^{I}$ should remain no dependence on the operator-insertion time $\tau $.}\BibitemShut {Stop}%
\bibitem [{\citenamefont {Sandvik}(1998)}]{Sandvik1998SAC}%
  \BibitemOpen
  \bibfield  {author} {\bibinfo {author} {\bibfnamefont {A.~W.}\ \bibnamefont {Sandvik}},\ }\bibfield  {title} {\bibinfo {title} {Stochastic method for analytic continuation of quantum monte carlo data},\ }\href {https://doi.org/10.1103/PhysRevB.57.10287} {\bibfield  {journal} {\bibinfo  {journal} {Phys. Rev. B}\ }\textbf {\bibinfo {volume} {57}},\ \bibinfo {pages} {10287} (\bibinfo {year} {1998})}\BibitemShut {NoStop}%
\bibitem [{\citenamefont {Sylju\aa{}sen}(2008)}]{OlavFSAC}%
  \BibitemOpen
  \bibfield  {author} {\bibinfo {author} {\bibfnamefont {O.~F.}\ \bibnamefont {Sylju\aa{}sen}},\ }\bibfield  {title} {\bibinfo {title} {Using the average spectrum method to extract dynamics from quantum monte carlo simulations},\ }\href {https://doi.org/10.1103/PhysRevB.78.174429} {\bibfield  {journal} {\bibinfo  {journal} {Phys. Rev. B}\ }\textbf {\bibinfo {volume} {78}},\ \bibinfo {pages} {174429} (\bibinfo {year} {2008})}\BibitemShut {NoStop}%
\bibitem [{\citenamefont {Sandvik}(2016)}]{Sandvik2016SAC}%
  \BibitemOpen
  \bibfield  {author} {\bibinfo {author} {\bibfnamefont {A.~W.}\ \bibnamefont {Sandvik}},\ }\bibfield  {title} {\bibinfo {title} {Constrained sampling method for analytic continuation},\ }\href {https://doi.org/10.1103/PhysRevE.94.063308} {\bibfield  {journal} {\bibinfo  {journal} {Phys. Rev. E}\ }\textbf {\bibinfo {volume} {94}},\ \bibinfo {pages} {063308} (\bibinfo {year} {2016})}\BibitemShut {NoStop}%
\bibitem [{\citenamefont {Shao}\ and\ \citenamefont {Sandvik}(2023)}]{HuiShaoSAC}%
  \BibitemOpen
  \bibfield  {author} {\bibinfo {author} {\bibfnamefont {H.}~\bibnamefont {Shao}}\ and\ \bibinfo {author} {\bibfnamefont {A.~W.}\ \bibnamefont {Sandvik}},\ }\bibfield  {title} {\bibinfo {title} {Progress on stochastic analytic continuation of quantum monte carlo data},\ }\href {https://doi.org/doi.org/10.1016/j.physrep.2022.11.002} {\bibfield  {journal} {\bibinfo  {journal} {Physics Reports}\ }\textbf {\bibinfo {volume} {1003}},\ \bibinfo {pages} {1} (\bibinfo {year} {2023})}\BibitemShut {NoStop}%
\bibitem [{\citenamefont {Yan}\ \emph {et~al.}(2021)\citenamefont {Yan}, \citenamefont {Wang}, \citenamefont {Ma}, \citenamefont {Qi},\ and\ \citenamefont {Meng}}]{yan2021topological}%
  \BibitemOpen
  \bibfield  {author} {\bibinfo {author} {\bibfnamefont {Z.}~\bibnamefont {Yan}}, \bibinfo {author} {\bibfnamefont {Y.-C.}\ \bibnamefont {Wang}}, \bibinfo {author} {\bibfnamefont {N.}~\bibnamefont {Ma}}, \bibinfo {author} {\bibfnamefont {Y.}~\bibnamefont {Qi}},\ and\ \bibinfo {author} {\bibfnamefont {Z.~Y.}\ \bibnamefont {Meng}},\ }\bibfield  {title} {\bibinfo {title} {Topological phase transition and single/multi anyon dynamics of $z_2$ spin liquid},\ }\href {https://doi.org/10.1038/s41535-021-00338-1} {\bibfield  {journal} {\bibinfo  {journal} {npj Quantum Materials}\ }\textbf {\bibinfo {volume} {6}},\ \bibinfo {pages} {1} (\bibinfo {year} {2021})}\BibitemShut {NoStop}%
\bibitem [{\citenamefont {Giamarchi}(2003)}]{Giamarchi2003}%
  \BibitemOpen
  \bibfield  {author} {\bibinfo {author} {\bibfnamefont {T.}~\bibnamefont {Giamarchi}},\ }\href {https://doi.org/10.1093/acprof:oso/9780198525004.001.0001} {\emph {\bibinfo {title} {Quantum Physics in One Dimension}}}\ (\bibinfo  {publisher} {Oxford University Press},\ \bibinfo {year} {2003})\BibitemShut {NoStop}%
\bibitem [{\citenamefont {Pereira}\ \emph {et~al.}(2006)\citenamefont {Pereira}, \citenamefont {Sirker}, \citenamefont {Caux}, \citenamefont {Hagemans}, \citenamefont {Maillet}, \citenamefont {White},\ and\ \citenamefont {Affleck}}]{Pereira2006}%
  \BibitemOpen
  \bibfield  {author} {\bibinfo {author} {\bibfnamefont {R.~G.}\ \bibnamefont {Pereira}}, \bibinfo {author} {\bibfnamefont {J.}~\bibnamefont {Sirker}}, \bibinfo {author} {\bibfnamefont {J.-S.}\ \bibnamefont {Caux}}, \bibinfo {author} {\bibfnamefont {R.}~\bibnamefont {Hagemans}}, \bibinfo {author} {\bibfnamefont {J.~M.}\ \bibnamefont {Maillet}}, \bibinfo {author} {\bibfnamefont {S.~R.}\ \bibnamefont {White}},\ and\ \bibinfo {author} {\bibfnamefont {I.}~\bibnamefont {Affleck}},\ }\bibfield  {title} {\bibinfo {title} {Dynamical spin structure factor for the anisotropic spin-$1/2$ heisenberg chain},\ }\href {https://doi.org/10.1103/PhysRevLett.96.257202} {\bibfield  {journal} {\bibinfo  {journal} {Phys. Rev. Lett.}\ }\textbf {\bibinfo {volume} {96}},\ \bibinfo {pages} {257202} (\bibinfo {year} {2006})}\BibitemShut {NoStop}%
\bibitem [{\citenamefont {Huang}\ \emph {et~al.}(2025)\citenamefont {Huang}, \citenamefont {Qi}, \citenamefont {Zhang},\ and\ \citenamefont {Lucas}}]{SWSSB_U1Hydrodynamics}%
  \BibitemOpen
  \bibfield  {author} {\bibinfo {author} {\bibfnamefont {X.}~\bibnamefont {Huang}}, \bibinfo {author} {\bibfnamefont {M.}~\bibnamefont {Qi}}, \bibinfo {author} {\bibfnamefont {J.-H.}\ \bibnamefont {Zhang}},\ and\ \bibinfo {author} {\bibfnamefont {A.}~\bibnamefont {Lucas}},\ }\bibfield  {title} {\bibinfo {title} {Hydrodynamics as the effective field theory of strong-to-weak spontaneous symmetry breaking},\ }\href {https://doi.org/10.1103/PhysRevB.111.125147} {\bibfield  {journal} {\bibinfo  {journal} {Phys. Rev. B}\ }\textbf {\bibinfo {volume} {111}},\ \bibinfo {pages} {125147} (\bibinfo {year} {2025})}\BibitemShut {NoStop}%
\bibitem [{\citenamefont {Guo}\ \emph {et~al.}(2025)\citenamefont {Guo}, \citenamefont {Yang},\ and\ \citenamefont {Yu}}]{yuxuejiaSWSSB}%
  \BibitemOpen
  \bibfield  {author} {\bibinfo {author} {\bibfnamefont {Y.}~\bibnamefont {Guo}}, \bibinfo {author} {\bibfnamefont {S.}~\bibnamefont {Yang}},\ and\ \bibinfo {author} {\bibfnamefont {X.-J.}\ \bibnamefont {Yu}},\ }\bibfield  {title} {\bibinfo {title} {Quantum strong-to-weak spontaneous symmetry breaking in decohered one-dimensional critical states},\ }\href {https://doi.org/10.1103/4vs5-l54f} {\bibfield  {journal} {\bibinfo  {journal} {PRX Quantum}\ }\textbf {\bibinfo {volume} {6}},\ \bibinfo {pages} {040311} (\bibinfo {year} {2025})}\BibitemShut {NoStop}%
\bibitem [{\citenamefont {Mermin}\ and\ \citenamefont {Wagner}(1966)}]{MWT1966}%
  \BibitemOpen
  \bibfield  {author} {\bibinfo {author} {\bibfnamefont {N.~D.}\ \bibnamefont {Mermin}}\ and\ \bibinfo {author} {\bibfnamefont {H.}~\bibnamefont {Wagner}},\ }\bibfield  {title} {\bibinfo {title} {Absence of ferromagnetism or antiferromagnetism in one- or two-dimensional isotropic heisenberg models},\ }\href {https://doi.org/10.1103/PhysRevLett.17.1133} {\bibfield  {journal} {\bibinfo  {journal} {Phys. Rev. Lett.}\ }\textbf {\bibinfo {volume} {17}},\ \bibinfo {pages} {1133} (\bibinfo {year} {1966})}\BibitemShut {NoStop}%
\bibitem [{\citenamefont {Hohenberg}(1967)}]{Hohenberg1967}%
  \BibitemOpen
  \bibfield  {author} {\bibinfo {author} {\bibfnamefont {P.~C.}\ \bibnamefont {Hohenberg}},\ }\bibfield  {title} {\bibinfo {title} {Existence of long-range order in one and two dimensions},\ }\href {https://doi.org/10.1103/PhysRev.158.383} {\bibfield  {journal} {\bibinfo  {journal} {Phys. Rev.}\ }\textbf {\bibinfo {volume} {158}},\ \bibinfo {pages} {383} (\bibinfo {year} {1967})}\BibitemShut {NoStop}%
\bibitem [{\citenamefont {Li}\ \emph {et~al.}(2024)\citenamefont {Li}, \citenamefont {Huang}, \citenamefont {Ding}, \citenamefont {Meng}, \citenamefont {Wang},\ and\ \citenamefont {Yan}}]{li2024relevant}%
  \BibitemOpen
  \bibfield  {author} {\bibinfo {author} {\bibfnamefont {C.}~\bibnamefont {Li}}, \bibinfo {author} {\bibfnamefont {R.-Z.}\ \bibnamefont {Huang}}, \bibinfo {author} {\bibfnamefont {Y.-M.}\ \bibnamefont {Ding}}, \bibinfo {author} {\bibfnamefont {Z.~Y.}\ \bibnamefont {Meng}}, \bibinfo {author} {\bibfnamefont {Y.-C.}\ \bibnamefont {Wang}},\ and\ \bibinfo {author} {\bibfnamefont {Z.}~\bibnamefont {Yan}},\ }\bibfield  {title} {\bibinfo {title} {Relevant long-range interaction of the entanglement hamiltonian emerges from a short-range gapped system},\ }\href {https://doi.org/10.1103/PhysRevB.109.195169} {\bibfield  {journal} {\bibinfo  {journal} {Phys. Rev. B}\ }\textbf {\bibinfo {volume} {109}},\ \bibinfo {pages} {195169} (\bibinfo {year} {2024})}\BibitemShut {NoStop}%
\bibitem [{\citenamefont {Wang}\ \emph {et~al.}(2025{\natexlab{b}})\citenamefont {Wang}, \citenamefont {Yang}, \citenamefont {Mao}, \citenamefont {Cheng},\ and\ \citenamefont {Yan}}]{wang2024sudden}%
  \BibitemOpen
  \bibfield  {author} {\bibinfo {author} {\bibfnamefont {Z.}~\bibnamefont {Wang}}, \bibinfo {author} {\bibfnamefont {S.}~\bibnamefont {Yang}}, \bibinfo {author} {\bibfnamefont {B.-B.}\ \bibnamefont {Mao}}, \bibinfo {author} {\bibfnamefont {M.}~\bibnamefont {Cheng}},\ and\ \bibinfo {author} {\bibfnamefont {Z.}~\bibnamefont {Yan}},\ }\bibfield  {title} {\bibinfo {title} {Sudden change in the entanglement hamiltonian: Phase diagram of an ising entanglement hamiltonian},\ }\href {https://doi.org/10.1103/ywk9-n2ds} {\bibfield  {journal} {\bibinfo  {journal} {Phys. Rev. B}\ }\textbf {\bibinfo {volume} {111}},\ \bibinfo {pages} {245126} (\bibinfo {year} {2025}{\natexlab{b}})}\BibitemShut {NoStop}%
\bibitem [{\citenamefont {Ding}\ and\ \citenamefont {Makivi\ifmmode~\acute{c}\else \'{c}\fi{}}(1990)}]{BKT034}%
  \BibitemOpen
  \bibfield  {author} {\bibinfo {author} {\bibfnamefont {H.-Q.}\ \bibnamefont {Ding}}\ and\ \bibinfo {author} {\bibfnamefont {M.~S.}\ \bibnamefont {Makivi\ifmmode~\acute{c}\else \'{c}\fi{}}},\ }\bibfield  {title} {\bibinfo {title} {Kosterlitz-thouless transition in the two-dimensional quantum xy model},\ }\href {https://doi.org/10.1103/PhysRevB.42.6827} {\bibfield  {journal} {\bibinfo  {journal} {Phys. Rev. B}\ }\textbf {\bibinfo {volume} {42}},\ \bibinfo {pages} {6827} (\bibinfo {year} {1990})}\BibitemShut {NoStop}%
\bibitem [{\citenamefont {Henelius}\ \emph {et~al.}(2002)\citenamefont {Henelius}, \citenamefont {Fr\"obrich}, \citenamefont {Kuntz}, \citenamefont {Timm},\ and\ \citenamefont {Jensen}}]{DL3_Henelius2002}%
  \BibitemOpen
  \bibfield  {author} {\bibinfo {author} {\bibfnamefont {P.}~\bibnamefont {Henelius}}, \bibinfo {author} {\bibfnamefont {P.}~\bibnamefont {Fr\"obrich}}, \bibinfo {author} {\bibfnamefont {P.~J.}\ \bibnamefont {Kuntz}}, \bibinfo {author} {\bibfnamefont {C.}~\bibnamefont {Timm}},\ and\ \bibinfo {author} {\bibfnamefont {P.~J.}\ \bibnamefont {Jensen}},\ }\bibfield  {title} {\bibinfo {title} {Quantum monte carlo simulation of thin magnetic films},\ }\href {https://doi.org/10.1103/PhysRevB.66.094407} {\bibfield  {journal} {\bibinfo  {journal} {Phys. Rev. B}\ }\textbf {\bibinfo {volume} {66}},\ \bibinfo {pages} {094407} (\bibinfo {year} {2002})}\BibitemShut {NoStop}%
\end{thebibliography}%

\clearpage

\appendix

\section{Appendix A: Stochastic Series Expansion and Directed-loop Algorithm}

Although the framework developed in the main text is applicable to both path-integral quantum Monte Carlo (QMC) and stochastic series expansion (SSE) QMC, we use the SSE formalism throughout this appendix because our implementation is built on SSE. We briefly review the SSE framework. In SSE, the partition function is expanded in powers of $-\beta H$:
\begin{equation}
Z =\mathrm{Tr} (e^{-\beta H}) = \sum_{\alpha} \sum_{S_M} \frac{\beta^n (M-n)!}{M!} \left\langle \alpha \left| \prod_{p=0}^{M-1} H_{a(p),b(p)} \right| \alpha \right\rangle
\end{equation}
where $\beta$ is the inverse temperature, and $S_M = \{[a(p),b(p)]\}_{p=0}^{M-1}$ is an operator string of fixed length $M$. The string contains $n$ non-identity operators and $M-n$ identity insertions. At each imaginary-time slice $p$, the operator $H_{a(p),b(p)}$ acts on a bond $b = \langle i,j \rangle$, with the index $a$ labelling the operator type. The initial state $|\alpha\rangle$ propagates through the operator string, and the sum over $|\alpha\rangle$ closes the trace.

Next, we follow the directed-loop algorithm~\cite{DL1_OlavF_2002,DL2_OlavF_2003}. Recall the standard XXZ Hamiltonian:
\begin{equation}
H = J\sum_{\langle i,j \rangle}\left(\Delta S^z_i S^z_j + \frac{1}{2}(S^+_i S^-_j+S^-_i S^+_j) \right) -h \sum_i S^z_i,
\label{eq:xxz_dl_h}
\end{equation}
the bond operators are split into diagonal and off-diagonal parts in SSE:
\begin{align}
H_{1,b} &= \left[C-\Delta S^z_i S^z_j+h_b(S^z_i+S^z_j)\right]_b, \nonumber\\
H_{2,b} &= \left[\frac{1}{2}(S^+_i S^-_j+S^-_i S^+_j)\right]_b,
\label{eq:dl_split_ops}
\end{align}
where $C=C_0+\varepsilon$, $C_0$ is chosen such that the diagonal matrix elements are already nonnegative, while the additional small parameter $\varepsilon>0$ is used to keep all diagonal matrix elements strictly positive. In the $S^z$ basis, one obtains six nonzero vertex types with weights
\begin{align}
\langle \uparrow \uparrow  |H_{1,b}|\uparrow \uparrow  \rangle  &= \varepsilon + 2h_b, \quad
\langle \downarrow \downarrow   |H_{1,b}|\downarrow \downarrow   \rangle = \varepsilon,\nonumber\\
\langle \uparrow \downarrow |H_{1,b}|\uparrow \downarrow \rangle  &=
\langle  \downarrow\uparrow |H_{1,b}| \downarrow \uparrow\rangle =\Delta/2+h_b +\varepsilon,\nonumber\\
\langle \downarrow \uparrow  |H_{2,b}|\uparrow \downarrow \rangle &=
\langle \uparrow \downarrow   |H_{2,b}|\downarrow \uparrow  \rangle = \frac{1}{2}.
\label{eq:dl_vertex_weights}
\end{align}
Here $h_b$ denotes the bond-distributed magnetic field. For a uniform field $h$, one usually assigns $h_b=h/(2d)$ on each bond connected to a site, where $d$ is the lattice coordination number. These six local weights define the directed-loop scattering events at each vertex. While in the present work we use the zero-field XXZ chain as the basic example throughout, the framework introduced here is not restricted to zero-field setting. 

In direct-loop practice, simulation sweeps alternate between diagonal updates and nonlocal (directed) loop updates. The truncation length $M$ is initialized such that $M \gtrsim \langle n\rangle_{\mathrm{eq}}\propto N\beta$, and is then increased during equilibration if $n$ approaches $M$ (e.g., $M \leftarrow \lfloor 4n/3 \rfloor$). After $M$ exceeds the typical expansion order by a sufficient margin, it is fixed for production measurements.

\section{Appendix B: Boundary-hole trick in Directed-loop Algorithm}

The directed-loop algorithm provides an efficient and general scheme for sampling the operator-string configuration space. In the loop update, the operator string is represented as a linked-vertex list, where each vertex records the local spin states on its legs and the operator type. The update line propagates a head through the linked-vertex list: starting from a chosen leg on a chosen vertex, the head enters a vertex and scatters to an exit leg according to a set of local probabilities, which flip the spin state on the traversed leg. The head then moves to the vertex connected via the exit leg and repeats the process, tracing a path --- the directed loop --- through the vertex network until it returns to its starting point. The key requirement is that the local scattering probabilities satisfy local detailed balance at every vertex~\cite{DL2_OlavF_2003}, expressed by
\begin{equation}
W_k\,P(e\vert i,k)=W_{k'}\,P(i\vert e,k'),
\label{eq:DL_local_balance}
\end{equation}
where $k$ is the vertex type (fully specifying the local spin configuration), $i$ and $e$ are the entrance and exit legs respectively, $P(e\vert i,k)$ is the probability of choosing exit $e$ given entrance $i$ at vertex $k$, and $k'$ is the vertex type after the leg flip. Eq.~(\ref{eq:DL_local_balance}) guarantees that each elementary scattering event is individually reversible, and that the composed directed-loop update satisfies detailed balance at the level of the full configuration. In what follows we explain why simple application of the directed-loop algorithm fails on the A-OBC/B-PBC manifold, and how the boundary-hole trick restores its validity at the open boundaries.

\begin{figure}[htb]
\centering
\includegraphics[width=\columnwidth]{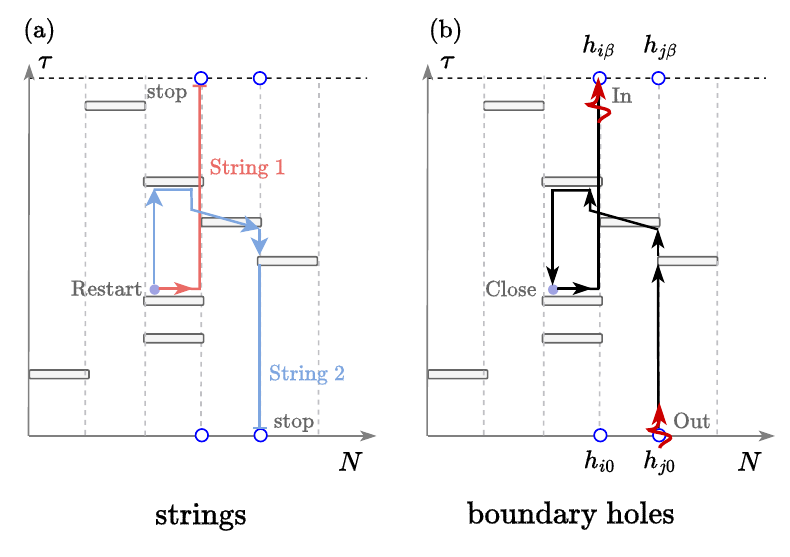}
\caption{Comparison between the two-string construction and the boundary-hole construction for directed-loop updates on the A-OBC/B-PBC manifold.
The gray bond operators denote local Hamiltonian terms distributed along both the spatial (horizontal) and imaginary-time (vertical) directions in the SSE operator-string configuration.
(a) In the two-string scheme, the loop head starts from a bulk leg and propagates by standard directed-loop scatterings. Upon reaching the open imaginary-time boundary, the first string (red) must terminate because no boundary scattering rule is available. Since the loop is not yet closed, one must restart from the original starting point and generate a second string (blue), which propagates until it also terminates at the boundary.
(b) In the boundary-hole construction, the open boundary endpoints are promoted to boundary holes $h_{i0},h_{j0},h_{i\beta},h_{j\beta}$, which serve as explicit entrance and exit channels for the loop head. The update is then completed through a hole-to-hole jump (red arrow) between boundary holes.} 
\label{fig: dlupdate}
\end{figure}

\paragraph{Why directed-loop strings fail.}
On the A-OBC/B-PBC manifold, worldlines in region $A$ have free endpoints at $\tau = 0$ and $\tau = \beta$. When the line head reaches such an endpoint, the standard directed-loop framework has no defined scattering rule. One natural but ultimately incorrect fix is to terminate the loop at the boundary, as illustrated in Fig.~\ref{fig: dlupdate}(a). In a standard construction, the loop starts from a randomly chosen leg on a bond vertex and then propagates through the linked list by local scatterings drawn from the directed-loop probabilities. It may encounter the situation marked as the red ``String~1'': upon hitting the open time boundary, there is no rule that guarantees the head can bypass the boundary and return to close the loop, and one is tempted to stop the construction there. However, at this stage the loop is not complete. The single red string leaves an unpaired link discontinuity at the starting point, which violates the conservation law~\cite{DL3_Henelius2002, DL2_OlavF_2003}. To repair this, one must restart from the original starting point and generate a second string (the blue ``String~2'') that carries the discontinuity through the bulk until it hits an open boundary again. After the second termination, the configuration contains no uncancelled discontinuity and one may regard the update as finished, implementing a global flip of the traversed paths. We refer to this two-string construction as the directed-loop \emph{string} update. Despite the fact that every bulk scattering event still satisfies the local directed-loop balance and has a well-defined reverse flow, the reverse move of the entire update is not guaranteed. The reason is the boundary termination: a true reverse process would require the in/out boundary events to be symmetric and to occur with equal probability, but the loop stop here provides no associated exit probability for launching the head with a prescribed incoming direction. Consequently, Eq.~(\ref{eq:DL_local_balance}) is violated precisely at the open boundaries, and detailed balance of the full update is then broken.

\begin{figure}[htb]
\centering
\includegraphics[width=\columnwidth]{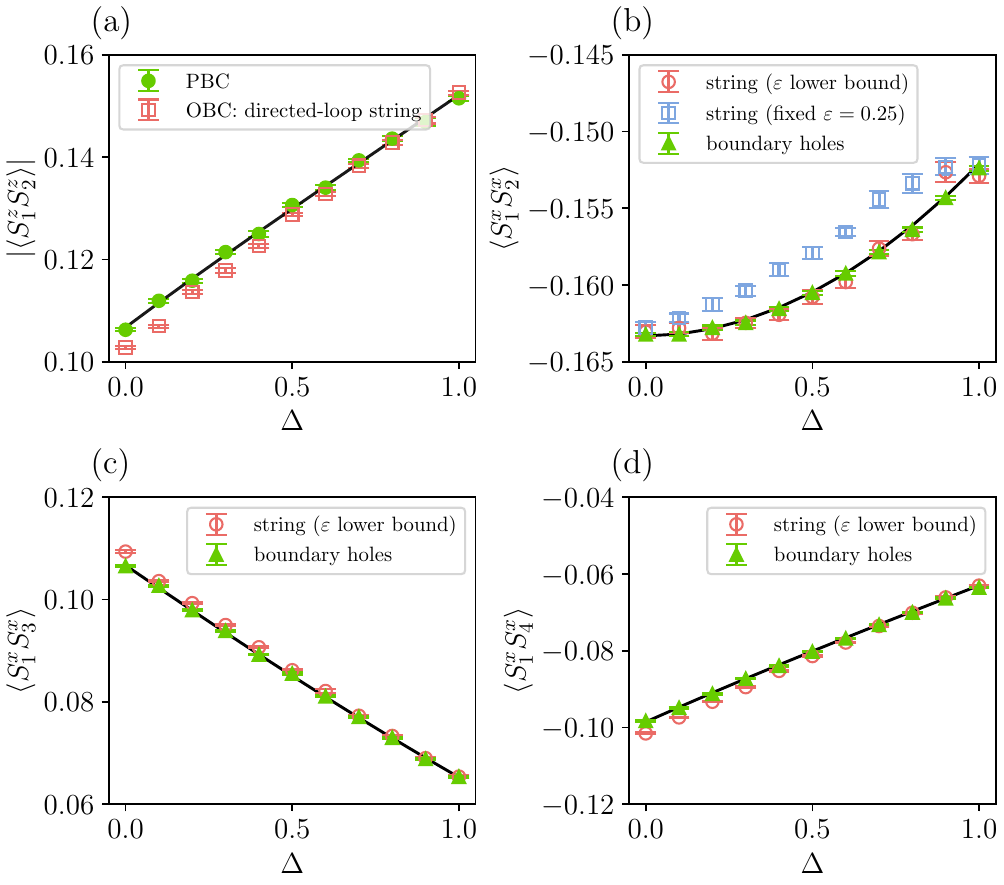}
\caption{Diagnosing and resolving the directed-loop drift on the A-OBC/B-PBC manifold.
(a) Benchmark of $\langle S_1^z S_2^z\rangle$ versus anisotropy $\Delta$ comparing ED (all solid black lines) with directed-loop measurements under imaginary-time PBC and the two-holes OBC (open imaginary-time boundaries at sites $i$ and $j$), demonstrating that the discrepancy originates from the OBC sector.
(b) For $\langle S_1^x S_2^x\rangle$, we compare ED with three directed-loop implementations: a fixed $\varepsilon=0.25$ choice (blue) and the lower-bound $\varepsilon$ (red) taken from the bounce-free construction of Refs.~\cite{DL1_OlavF_2002}, both used directed-loop string update, and compare the correct results of directed-loop update using boundary-hole trick (green).
(c)-(d) The remaining drift comes from DL string method is exposed by longer-range correlators $\langle S_1^x S_3^x\rangle$ and $\langle S_1^x S_4^x\rangle$.
In contrast, the boundary-hole trick (green) removes the drift in (b)-(d) and restores agreement with ED across $\Delta$.
} 
\label{fig: dlcompare}
\end{figure}

As shown in Fig.~\ref{fig: dlcompare}, this violation appears as a systematic drift in benchmark observables relative to exact diagonalization (ED) as anisotropy $\Delta$ increases. More troublingly, the drift magnitude depends on the parameter $\varepsilon$, which appears in the vertex weights of Eq.~\eqref{eq:dl_vertex_weights}. Since $\varepsilon$ controls the weights of part of the operators, this indicates that the probabilities of these operators in the loop update do not satisfy detailed balance. In Fig.~\ref{fig: dlcompare}(a), we compare
ED, the directed-loop algorithm under imaginary-time PBC, and the string update (labeled as ``string'') under two-site OBC, i.e., with open imaginary-time boundaries at sites $i$ and $j$, showing that the problem originates from OBC.   In Fig.~\ref{fig: dlcompare}(b), we measure different string cases that violate detailed balance, one with fixed $\varepsilon=0.25$ and the other with the lower-bound $\varepsilon = \frac{1-\Delta}{4}$ given by the bounce-free solution in Ref.~\cite{DL1_OlavF_2002}. The latter looks better, but in Fig.~\ref{fig: dlcompare}(c)-(d) this drift is still exposed.

\paragraph{Boundary-hole construction.}
The core challenge is to satisfy local detailed balance while respecting the rules introduced by OBC. Therefore, we restore a valid closed-loop topology by introducing \emph{boundary holes} that represent the open imaginary-time endpoints of region $A$ as explicit scattering channels in the vertex network. The open boundary of region $A$ has two edges in imaginary time: the lower edge at $\tau = 0$ and the upper edge at $\tau = \beta$. For each site $i \in A$, the corresponding boundary holes at these two edges are:
\begin{equation}
h_{i0}\equiv (i,\tau=0^-),\qquad
h_{i\beta}\equiv (i,\tau=\beta^+).
\end{equation}
Each boundary hole carries a single leg (the worldline endpoint)  and is included in a common connected hole pool together with all other active holes in the configuration. When the line head enters a hole $h$,  it performs a \textbf{hole-to-hole jump} to an exit hole $h'$ drawn from the pool. The key requirement on this jump is that it satisfies the same local detailed balance as Eq.~\eqref{eq:DL_local_balance}.  Since holes carry no internal vertex weight, the local balance condition reduces to the symmetry condition with probability:
\begin{equation}
P (h \to h') = P(h' \to h)
\label{eq:teleport_rule}
\end{equation}
Any symmetric jump probability satisfies Eq.~\eqref{eq:teleport_rule}. For simplicity we choose the uniform distribution over all holes in the pool, including $h$ itself (i.e., the line head may exit from the same hole it entered). In this simple way, $P(h\rightarrow h')=\frac{1}{N_h}$ for any $h'$. $N_h$ is the total number of holes currently in the pool. After the jump, the line head exits from $h'$ and continues to scatter through the bulk using the standard directed-loop rules of Eq.~\eqref{eq:DL_local_balance}. The boundary-hole jump thereby converts the otherwise open trajectory into a closed path on the extended vertex graph that includes the hole pool.

The benchmark results from the boundary-hole trick are shown in Fig.~\ref{fig: dlcompare}(b)-(d): no drift with increasing $\Delta$ is observed, and the agreement with ED is good. As the final words of this section, the complete directed-loop update on the A-OBC/B-PBC manifold consists of two types of elementary events: (i) bulk vertex scatterings, each of which satisfies Eq.~\eqref{eq:DL_local_balance} by the standard directed-loop construction, and (ii) hole-to-hole jumps, each of which satisfies the symmetric condition Eq.~\eqref{eq:teleport_rule}. The same boundary-hole construction extends straightforwardly to cluster-like updates: boundary holes serve as scattering channels that reconnect the open cluster boundary into a closed cluster, and the jump rule ensures detailed balance in that context as well.

\section{Appendix C: GRDM method and off-diagonal correction function}

We now provide the technical details of the GRDM operator insertion and the corresponding Monte Carlo estimator.
Recall from the main text that the generalized reduced density operator is defined by inserting $\hat O_A$ at imaginary time $\tau$ into the A-OBC/B-PBC manifold,
\begin{equation}
{\rho}_A^{\hat{O}}(\tau) = \frac{\mathrm{Tr}_B \left(e^{-(\beta-\tau) H} \hat{O}_A e^{-\tau H}\right)}{Z}
\label{rhoaoAppendix}
\end{equation}
where $Z$ is the standard normalized factor which $Z=\mathrm{Tr} \left(e^{-\beta H} \right)$. Consider an insertion of $\hat O_A$ at position $\tau$. Let $\tilde{\alpha}_A$ and $\tilde{\alpha}'_A$ be the propagated $A$-states immediately before and after position $\tau$, respectively. The corresponding SSE configuration weight is
\begin{align}
\widetilde W_{\hat O}(\alpha_A,\alpha_A';\{S_n\},\tau)
&=
W(\alpha_A,\tilde{\alpha}_A;\{S_{\tau'<\tau}\}) \times \nonumber\\
&\quad
\langle \tilde{\alpha}_A | \hat O_A | \tilde{\alpha}'_A \rangle \times
W(\tilde{\alpha}'_A,\alpha_A';\{S_{\tau'>\tau}\})
\label{eq:Wtilde_O}
\end{align}
which is the operator-string representation of Eq.~(\ref{rhoaoAppendix}). The propagation is split into two segments by the insertion at $\tau$, and the factor
$\langle \tilde{\alpha}_A | \hat O_A | \tilde{\alpha}'_A \rangle$
is precisely the matrix element of the inserted operator between the propagated states immediately below and above the insertion.
For a diagonal operator, $ |\tilde{\alpha}_A \rangle = | \tilde{\alpha}'_A \rangle$ and this reduces to an eigenvalue. For an off-diagonal operator such as $\sigma^x_i$, the factor is non-zero only when $|\tilde{\alpha}_A \rangle$ differs from $| \tilde{\alpha}'_A \rangle$ by a single spin flip at site $i$.

\paragraph{Joint sampling and ratio estimator.}
Because an insertion changes the configuration weight, we evaluate normalized observables via a joint Markov chain that alternates between
(i) an identity-inserted sector $\widetilde W_{\hat I}$ and
(ii) an operator-inserted sector $\widetilde W_{\hat O}$ at the same $\tau$.
A Metropolis switch $\hat O \leftrightarrow \hat I$ is defined as
\begin{equation}
P(\hat O\rightarrow \hat I)=\min\!\left(1,\frac{\widetilde W_{\hat I}}{\widetilde W_{\hat O}}\right),
\quad
P(\hat I\rightarrow \hat O)=\min\!\left(1,\frac{\widetilde W_{\hat O}}{\widetilde W_{\hat I}}\right)
\label{eq:sector_switch}
\end{equation}
where $P$ denotes the acceptance probability. For any operator $\hat O^{(1)}_A$ supported in region $A$, the corresponding imaginary-time correlator is obtained as a ratio of Monte Carlo averages over the combined chain,
\begin{equation}
\langle \hat O_1(\tau)\,\hat O_2(0)\rangle
=
\frac{\mathrm{Tr}\bigl[\tilde{\rho}_A^{\hat{O}^{(1)}}(\tau)\,\hat O^{(2)}_A\bigr]}
{ \mathrm{Tr}\bigl[\tilde{\rho}_A^{\hat I}\bigr]}
\label{eq:ratio_estimator}
\end{equation}
where $\tilde{\rho}_A^{\hat O^{(1)}}$ and $\tilde{\rho}_A^{\hat I}$ denote the generally unnormalized matrices associated with $\widetilde W_{\hat O^{(1)}}$ and $\widetilde W_{\hat I}$, respectively. Here $\hat O_A^{(2)}$ is another arbitrary local operator supported in $A$, while the SSE-GRDM simulation directly measures the operator-inserted reduced density matrix $\rho_{\hat O^{(1)}}(\tau)$. The ratio form guarantees the correct normalization, and the imaginary-time correlator is then obtained in post-processing by contracting the measured matrix with $\hat O_A^{(2)}$, as in Eq.~(\ref{eq:ratio_estimator}).

\begin{figure}[htb]
\centering
\includegraphics[width=\columnwidth]{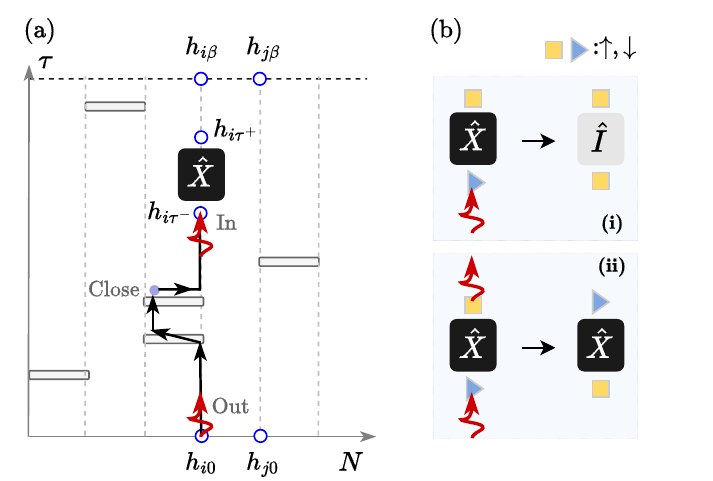}
\caption{Directed-loop treatment of an $X$ insertion using operator holes.
(a) Schematic configuration on the A-OBC/B-PBC manifold with an inserted operator $\hat X$ at site $i$ and imaginary-time position $\tau$. The forced stop at the insertion is resolved by introducing two internal operator holes, $h_{i\tau^-}$ and $h_{i\tau^+}$, located immediately below and above the insertion, respectively. These operator holes are treated in the same way as the boundary holes $h_{i0}, h_{j0}, h_{i\beta}, h_{j\beta}$. The configuration shown here illustrates one specific case, in which the update is completed through an operator-hole to boundary-hole jump (red arrow).
(b) Blue triangles and yellow squares denote the two local spin states, $|\uparrow \rangle$ and $|\downarrow \rangle$, respectively. Two possible cases of the hole-jump rule at the insertion. In case (i), the jump leads to the switch $\hat X\rightarrow \hat I$. In case (ii), the line head enters from one operator hole and exits from the opposite hole, which is equivalent to passing straight through the insertion, so the operator type remains $\hat X$. Boundary-hole to boundary-hole jumps obey the same rule, although they are not shown explicitly here.} 
\label{fig:Xswitch_holes}
\end{figure}

\paragraph{$X$ insertion.}
As a concrete illustration, consider the computation of the off-diagonal correlation function in the $S^z$ basis. Let the inserted operator be $\hat O_A=\hat X_i$, where $X$ may be viewed as either $\sigma^x$ or $S^x$ in the main text. For convenience, in this section we take $X=\sigma^x$. The relevant matrix elements are:
\begin{equation}
\langle \uparrow | \hat{X} | \downarrow \rangle = 1, \quad
\langle \downarrow | \hat{X} | \uparrow \rangle = 1, \quad
\langle s | \hat{I} | s \rangle = 1 \quad (s = \uparrow, \downarrow)
\end{equation}
The $\hat X \leftrightarrow \hat I$ switch involves an update event of the stop type. When the line head reaches the insertion, it is forced to stop there, and this stop triggers a change of operator type between the off-diagonal $\hat X$ and the diagonal identity $\hat I$. However, because such a stop cuts the worldline at the insertion site in region $A$, it introduces a local discontinuity in the loop construction. To restore a valid directed-loop update again, we introduce an operator-hole pair, namely two internal holes $h_{i\tau^-}$ and $h_{i\tau^+}$ located immediately below and above the insertion time slice $\tau$, respectively, as illustrated in Fig.~\ref{fig:Xswitch_holes}(a). These two internal holes are incorporated into the same hole pool as the boundary holes introduced in Appendix~B.

Once the line head reaches one of these holes, it performs the same symmetric hole-to-hole jump as for the boundary holes, in accordance with Eq.~(\ref{eq:teleport_rule}). Figure~\ref{fig:Xswitch_holes}(a) illustrates one specific possibility, namely an operator-hole to boundary-hole jump.  This can be viewed as a ``stop'',
since the single spin along the path is flipped, the operator $\hat X$ changes its type to the identity operator $\hat{I}$, as indicated in Fig.~\ref{fig:Xswitch_holes}(b) $\rm(i)$ case.
However, the operator-hole to boundary-hole jump is not the only allowed case. There is Fig.~\ref{fig:Xswitch_holes}(b) $\rm(ii)$ case, the line head may also jump from one operator hole to the other, i.e., from $h_{i\tau^-}$ to $h_{i\tau^+}$ or vice versa. This process corresponds to passing straight through the insertion, so the operator type remains unchanged. Boundary-hole to boundary-hole jumps are also part of the same unified hole-jump rule, although that case is not shown explicitly in the figure.

Therefore, the operator-hole construction provides a unified framework in which boundary holes and insertion-induced holes are treated on the same footing, and different hole-to-hole connections naturally encode either sector switching or propagation within the same sector. The loop update procedure is summarized in Algorithm~\ref{alg:loop_update_sigmaX}. For clarity, the hole-jump processes are grouped into three classes: BHBH (boundary-hole to boundary-hole), XHBH (operator-hole to boundary-hole), and XHXH (operator-hole to operator-hole).

Finally, we note that the imaginary-time argument $\tau$ of the correlator corresponds to the actual SSE position index $p_\tau$ in the operator string. Provided that the SSE truncation $M$ is sufficiently large compared with the saturation value reached during thermalization, we may accurately employ the linear mapping~\cite{Sandvik2010Computational, BRAZhiyan}:
\begin{equation}
  \tau = p_\tau \frac{\beta}{M}
  \label{eq:tautop}
\end{equation}
$\beta$ is the inverse temperature. The imaginary time $\tau$ is interpreted, in an average sense, through the mapping in Eq.~\eqref{eq:tautop} when evaluating imaginary-time correlation functions. Therefore, we mainly use $\tau$ instead of $p_\tau$ to refer to the insertion operator information. Benchmark results for this approximation are presented in the main text Fig.~\ref{fig:imcorbenchmark}.
Since the simulation protocol requires an operator insertion at a prescribed $p_\tau$, we generally fix $M$ to be a sufficiently large constant. 

Furthermore, this framework is not restricted to imaginary-time correlations. Equal-time observables are obtained simultaneously. By isolating and treating the $\tilde{\rho}_{\hat{I}}$ component from Eq.~\eqref{eq:ratio_estimator} independently, we obtain: 
\begin{equation}
  \langle \hat O^{(1)}_A\,\hat O^{(2)}_A\rangle
= \frac{\mathrm{Tr}\bigl[\tilde{\rho}_A^{\hat I}\,O^{(1)}_A\hat O^{(2)}_A\bigr]} { \mathrm{Tr}\bigl[\tilde{\rho}_A^{\hat I}\bigr]}
\label{eq:GRDM-eqcorr}
\end{equation}
This expression is consistent with the correlation formula in Eq.~\eqref{eq:ratio_estimator} for the specific case where $\hat{O}^{(1)}_A$ is positioned at $p_\tau = 0$. We also present equal-time correlation data in Fig.~\ref{fig:imcorbenchmark} (b) of the main text. There, the horizontal axis $\tau$ merely labels the imaginary-time position at which the GRDM insertion operator is placed, whereas the $\rho_{\hat I}$ is used to evaluate the equal-time correlation. In other words, while computing imaginary-time correlations, we simultaneously obtain a collection of RDMs that can also be used to evaluate static observables, such as the equal-time correlation function shown in Eq.~\eqref{eq:GRDM-eqcorr}.

\begin{algorithm}[t]
\caption{Loop update with boundary holes and $X$ insertion}
\label{alg:loop_update_sigmaX}
\KwIn{An SSE operator-string configuration $\{H_{t,b}\}$ with a single-site $X/I$ insertion at \texttt{insert(1)}, together with the boundary-hole locations $(\texttt{open\_i},\texttt{open\_j})$.}
\KwOut{An updated operator string and spin configuration.}

Determine the allowed local update channels from the current spin states\;
\For{a prescribed number of loop attempts}{
  Choose a random allowable leg $v_0$ as the starting point and set $v_1\leftarrow v_0$\;

  \Repeat{$v_1$ returns to $v_0$}{
    Identify the current vertex / insertion position from $v_1$\;

    \uIf{$v_1$ hits the single-site insertion \texttt{insert(1)}}{
      \tcp{operator-hole sector}
      \uIf{the jump is of type XHBH, with probability \texttt{pXHBH}}{
        switch $X \leftrightarrow I$\;
        relocate the loop head from an operator hole to a boundary hole\;
      }
      \Else{
        \tcp{XHXH process}
        pass through the insertion by jumping between the two operator holes\;
        keep the operator type unchanged\;
      }
      advance the loop head to the next linked leg\;
      \textbf{continue}\;
    }
    \Else{
      \tcp{bulk two-body vertex}
      apply the standard directed-loop scattering rule\;
    }

    \If{the propagated leg reaches a boundary-hole site}{
      \uIf{the jump is of type BHBH, with probability \texttt{pBHBH}}{
        relocate the loop head to a boundary hole\;
      }
      \Else{
        \tcp{boundary hole to insertion sector}
        relocate the loop head to one of the two operator holes, with probability $1/2$ each\;
      }
    }
    \Else{
      continue the usual linked-vertex propagation\;
    }
  }
}
\end{algorithm}

\section{Appendix D: Simulating the R\'enyi-1 Correlator in Thermal Gibbs States}

In this appendix, we provide the detail for measuring the R\'enyi-1 correlator $R_1(i,j)$ in the total $S^z$-conserving XXZ model.

In canonical purification, a mixed state $\rho$ is mapped to
$|\sqrt{\rho}\rangle\rangle$ in a doubled Hilbert space and
the R\'enyi-1 correlator probes correlations between the
left and right sectors of this purified state~\cite{SWSSBZack}. For a symmetry-charged local
operator $O_i$, one considers the composite operator
$O_{ij}=O_i O_j^\dagger$ and the corresponding doubled-space correlator
$R_1(i,j)=\langle\langle \sqrt{\rho}|\, O^{L}_{ij}\,\bar{O}^{R}_{ij}\,|\sqrt{\rho}\rangle\rangle$, 
where $L/R$ label the two copies, and the bar denotes the natural conjugation
action on the right copy. \ When $\rho$ respects a global symmetry, the purified state can preserve a strong symmetry $U(L)\otimes \bar{U}(R)$ in the doubled space, which forces $R_1$ to vanish. As temperature or other control parameters are tuned, the onset of strong-to-weak spontaneous symmetry breaking (SWSSB) reduces this strong symmetry to its diagonal subgroup. This gives rise to long-range order in the R\'enyi-1 correlator, with $R_1(i,j)\to c>0$ as $|i-j|\to\infty$.  However, precisely because the strong symmetry is only broken down to the weak symmetry, the standard two-point correlation function does not develop
long-range order and still decays to zero in the thermodynamic limit. Therefore,
an SWSSB phase can be identified by the coexistence of a vanishing conventional
two-point correlator and long-range order in the R\'enyi-1 correlator.

For the $U(1)$-symmetric XXZ model, the conserved charge is the total magnetization $Q \equiv \sum_{i} S_i^z$, 
hence a natural choice of symmetry-charged local operator is $O_i=S_i^{+}$ (or $S_i^{-}$). 
This is because $[Q,S_i^{+}] = S_i^{+}$, i.e., $S_i^{+}$ carries $U(1)$ charge $+1$ (and similarly $[Q,S_i^-]=-S_i^-$).
With this choice, the R\'enyi-1 correlator can be written in operator form as
\begin{equation}
 R_1(i,j) =  \mathrm{Tr} \left(
    \sqrt{\rho}\, S_i^+ S_j^- \,
    \sqrt{\rho}\, S_i^- S_j^+
  \right)
  \label{eq:R1_sqrt_rho_def}
\end{equation}

However, directly measuring $\sqrt{\rho}$ is not scalable because it depends on $\rho$ which is still strongly limited by system size. We use the thermal Gibbs state $\rho_\beta=e^{-\beta H}/Z$~\cite{SWSSBZack}.
Substituting $\sqrt{\rho_\beta}=e^{-\beta H/2}/\sqrt{Z}$ into Eq.~\eqref{eq:R1_sqrt_rho_def} yields,
\begin{align}
  R_1(i,j)
  &=
  \frac{1}{Z}\,
  \mathrm{Tr}\!\left[
    e^{-\beta H/2}\, S_i^+ S_j^-\,
    e^{-\beta H/2}\, S_i^- S_j^+
  \right]
  \nonumber\\
  &=
  \frac{1}{Z}\,
  \mathrm{Tr} \left[
    e^{-\beta H}\,
    \left(e^{+\beta H/2} S_i^+ S_j^- e^{-\beta H/2}\right)\,
    S_i^- S_j^+
  \right]
  \nonumber\\
  &\equiv
  \left\langle
    S_i^+ S_j^-(\beta/2)\,S_i^- S_j^+
  \right\rangle
  \label{eq:R1_tau_form_3}
\end{align}
Eq.~\eqref{eq:R1_tau_form_3} is the key step that makes large-scale QMC evaluation possible: $R_1$ becomes an imaginary-time-ordered four-point correlation function,
with the charge-neutral composite operator $O_{ij}=S_i^+S_j^-$ placed at $\tau=\beta/2$ and its conjugate $O_{ij}^\dagger=S_i^-S_j^+$ at $\tau=0$.

To diagnose SWSSB, the R\'enyi-1 correlator is evaluated in a fixed charge sector. Let $P_q$ be the projector onto the subspace with total charge $Q=q$ (in this work we focus on $q=0$). The corresponding projected thermal state is
\begin{equation}
  \rho_{q}
  \equiv
  \frac{P_q\,e^{-\beta H}}{\mathrm{Tr}\left(P_q\,e^{-\beta H}\right)} 
  \label{eq:rho_projected}
\end{equation}
which is simply the Gibbs ensemble restricted to the fixed-$Q$ sector.
Accordingly, the fix-sector R\'enyi-1 correlator is defined by replacing $Z \mapsto Z_q$ and restricting the ensemble average in Eq.~\eqref{eq:R1_tau_form_3} to $\langle \cdots \rangle_q$, where $Z_q = \mathrm{Tr}(P_q e^{-\beta H})$.

We evaluate $R_1$ using the GRDM insertion at $\tau=\beta/2$.
The composite operator $O_{ij}=S_i^+S_j^-$ is inserted, and the conjugate $O_{ij}^\dagger=S_i^-S_j^+$ is measured by post-processing matrix multiplication. 
Following the operator-switch construction introduced in Appendix~C, and applying it here to the analogous $S_i^+ S_j^- \leftrightarrow I_i I_j$ switch, the R\'enyi-1 correlator is evaluated as the ratio estimator:
\begin{equation}
  R_1(i,j)
  =
  \frac{
    \mathrm{Tr} \left[\rho_A^{S_i^+ S_j^-}(\tau=\beta/2)\, S_i^- S_j^+\right]_q
  }{
    \mathrm{Tr}\left[\rho_{\hat I}\right]_q
  }
  \label{eq:R1_ratio_estimator}
\end{equation}
here the same Markov chain accumulates the numerator and denominator contributions under the fixed total $S^z$ constraint.  From a technical standpoint, the switch $S_i^{+}S_j^{-}\leftrightarrow I_i I_j$ only requires solving about half of the directed-loop equations compared to the standard case. The reason is that for $j\neq i+1$ one may treat $S_i^{+}S_j^{-}$ as a long-distance two-body operator, while $I_i I_j$ does not change any spin state and can be regarded as a two-body operator with unit weight. Therefore, the update involves only five relevant vertices, i.e., the long-distance $S_i^{-}S_j^{+}$-type vertex is not included.

\begin{figure}[htb]
\centering
\includegraphics[width=\columnwidth]{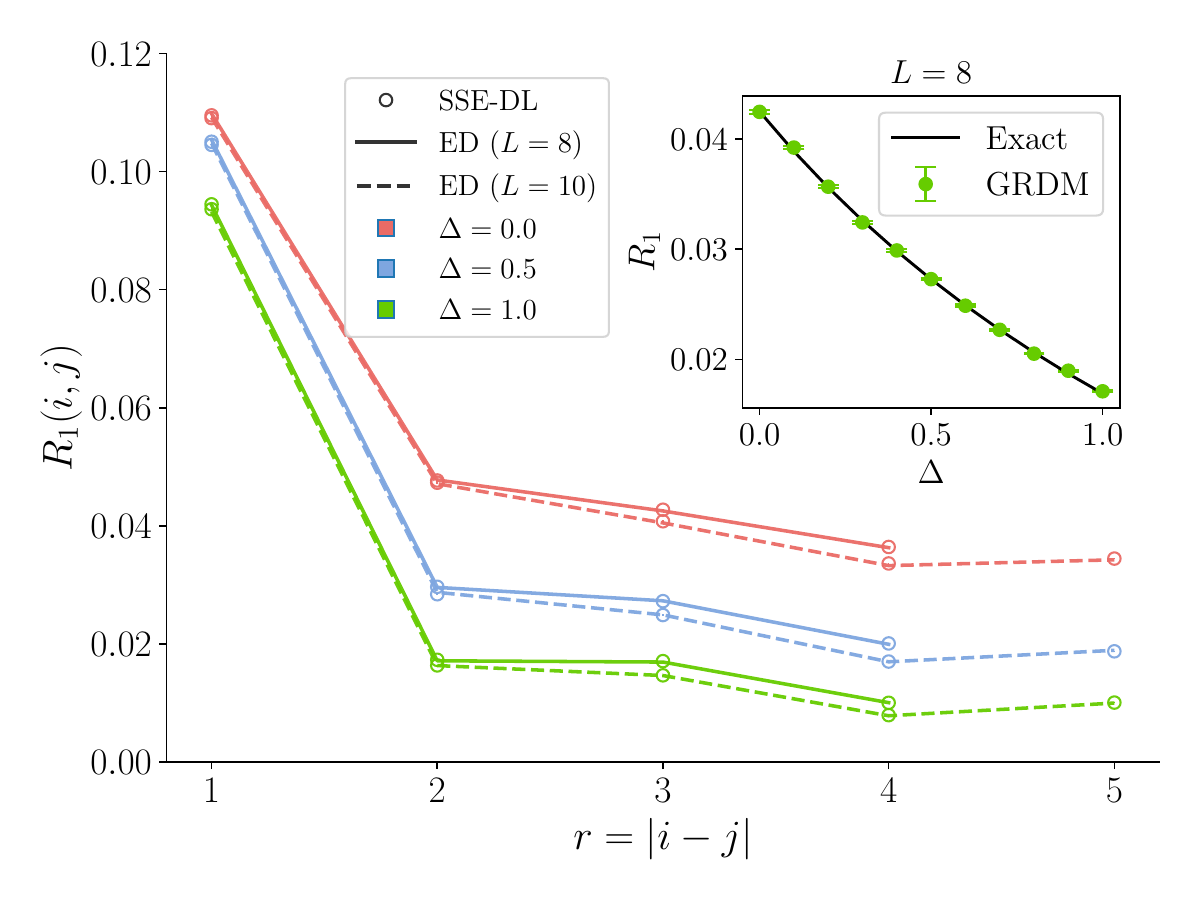}
\caption{Benchmark of the R\'enyi-1 correlator $R_1(i,j)$ in the 1D XXZ chain.
Main panel: $R_1$ versus separation $r=|i-j|$ for system sizes $L=8$ (solid lines) and $L=10$ (dashed lines), at several anisotropies $\Delta$ (color coded). Symbols show SSE directed-loop (SSE-DL) measurements by GRDM framework, while lines are ED results; the agreement within error bars validates the GRDM-based estimator.
Inset: $R_1$ as a function of $\Delta$ for $L=8$, comparing SSE-DL data with the exact curve from ED.} 
\label{fig:R1benchmark}
\end{figure}

We benchmark the implementation against ED on periodic XXZ chains. Fig.~\ref{fig:R1benchmark} shows $R_1(r)$ with $r=|i-j|$ at several anisotropies $\Delta$ (color coded), comparing GRDM/SSE-DL measurements (symbols) against ED results (lines) for $L=8$ and $L=10$. Across the entire distance range shown, the QMC data reproduce the ED results within error bars for both system sizes, demonstrating that the GRDM insertion together with the fix-sector constraint yields an unbiased estimator of $R_1$. The inset further shows $R_1$ as a function of $\Delta$ at fixed $L=8$, where the GRDM data agree with the exact curve over the full anisotropy range, providing an additional consistency check of the implementation.  For the larger-system results shown in the main text, we focus on the longest-distance correlator $R_1(r=L/2)$ and simulating system sizes $L=\{20,24,28,32\}$, discarding smaller sizes $L=\{8,12,16\}$ to reduce the visible finite-size drift. In that analysis, the data are fitted to a constant $R_1 = c$, 
so as to test whether the long-distance correlator approaches a nonzero plateau in the accessible size window.

\end{document}